\documentclass[10pt,preprint,flushrt]{aastex}
\slugcomment{{\em Astrophys.\  J.,} v.\  634, in press (November 20, 2005)}

\shorttitle{Threshold Rotation Rates for Be Star Disks}
\shortauthors{Steven R. Cranmer}

\begin{document}

\title{A Statistical Study of Threshold Rotation Rates for
the Formation of Disks around Be Stars}

\author{Steven R. Cranmer}
\affil{Harvard-Smithsonian Center for Astrophysics,
60 Garden Street, Cambridge, MA 02138}
\email{scranmer@cfa.harvard.edu}

\begin{abstract}
\baselineskip=11.7pt
This paper presents a detailed statistical determination of
the equatorial rotation rates of classical Be stars.
The rapid rotation of Be stars is likely to be linked to
the ejection of gas that forms dense circumstellar disks.
The physical origins of these disks are not understood, though
it is generally believed that the ability to spin up matter
into a Keplerian disk depends on how close the stellar rotation
speed is to the critical speed at which the centrifugal force
cancels gravity.
There has been recent disagreement between the traditional
idea that Be stars rotate between 50\% and 80\% of their
critical speeds and new ideas (inspired by the tendency for
gravity darkening to mask rapid rotation at the equator) that
their rotation may be very nearly critical.
This paper utilizes Monte Carlo forward modeling to
simulate distributions of the projected rotation speed
($v \sin i$), taking into account gravity darkening,
limb darkening, and observational uncertainties.
A chi-squared minimization procedure was used to find the
distribution parameters that best reproduce observed
$v \sin i$ distributions from R.\  Yudin's database.
Early-type (O7e--B2e) Be stars were found to exhibit a
roughly uniform spread of intrinsic rotation speed that
extends from 40--60\% up to 100\% of critical.
Late-type (B3e--A0e) Be stars exhibit progressively narrower
ranges of rotation speed as the effective temperature
decreases; the lower limit rises to reach critical rotation
for the coolest Be stars.
The derived lower limits on equatorial rotation speed
represent conservative threshold rotation rates for the onset
of the Be phenomenon.
The significantly subcritical speeds found for early-type Be
stars represent strong constraints on physical models of
angular momentum deposition in Be star disks.
\end{abstract}


\section{Introduction}

Be stars are rapidly rotating, non-supergiant B-type stars that
exhibit, or have exhibited in the past, emission in their
hydrogen Balmer lines.
The observed properties of Be stars are consistent with the
coexistence of a dense circumstellar disk (flattened in the
plane perpendicular to the rotation axis) and a variable stellar
wind (Struve 1931; Doazan 1982; Slettebak 1988; Prinja 1989;
Porter \& Rivinius 2003).
The gas in the so-called ``decretion disk'' is traditionally
believed to be ejected from the star and not accreted from
an external source (see, however, Harmanec et al.\  2002;
Abt 2004).
Although there is increasing evidence that the disk gas is in
Keplerian orbit (e.g., Hanuschik 1996; Hummel \& Vrancken 2000),
there is a great deal of evidence that Be-star photospheres are
rotating too {\em slowly} to propel any atmospheric material
into orbit.
Typical observationally determined values of the ratio of
equatorial rotation speed to the critical rotation speed
$V_{\rm crit}$ (at which gravity is balanced by outward
centrifugal forces) range between 0.5 and 0.8 (Slettebak 1982;
Porter 1996; Yudin 2001).
If this is the case, then any theoretical model for the origin
of Be-star disks would require a substantial increase in
angular momentum between the photosphere and the inner edge of
the disk.

Recently, the idea that Be stars are rotating with significantly
subcritical rotation speeds has been called into question.
The primary observational diagnostic of hot-star rotation is
the Doppler broadening of photospheric absorption lines,
first elucidated by Abney (1877).
Rotational broadening provides a surface-weighted measure of
the product of the equatorial rotation speed $V_{\rm eq}$ and
$\sin i$, where $i$ is the inclination angle to the observer.
Traditional means of determining $V_{\rm eq} \sin i$ from
line profiles (e.g., Tassoul 1978; Gray 1992) often assume
that the star is spherical.
However, rapidly rotating O and B stars tend to become
centrifugally distorted into oblate shapes and thus undergo
``gravity darkening'' (i.e., a redistribution of radiative
flux in proportion to the centrifugally modified gravity;
von Zeipel 1924).
The equators of such distorted stars become dimmer and cooler
than their poles, and thus the most rapidly rotating regions
of the stellar surface are weighted less strongly in the
resulting star-averaged absorption profiles.
This tendency for gravity-darkened stars to exhibit narrower
profiles than would be the case for spherical stars---and
thus lower computed values of $V_{\rm eq} \sin i$---has been
known for more than a half century (e.g., Slettebak 1949;
Stoeckley 1968; Hardorp \& Strittmatter 1968;
Walker et al.\  1979; Collins \& Truax 1995) and has been
recently highlighted as a potential bias in statistical
samples of Be star rotation rates (Zorec et al.\  2003;
Townsend et al.\  2004; Cohen et al.\  2005;
Fr\'{e}mat et al.\  2005).

A physical understanding of the Be phenomenon hinges on how
close the stars are rotating to their critical speeds.
If $V_{\rm eq}$ is within one or two sound speeds of
$V_{\rm crit}$ (which would imply $V_{\rm eq}/V_{\rm crit}
\gtrsim 0.95$), there are many possible weak processes
that could easily propel gas into orbit (see, e.g.,
Owocki 2005).
When the above ratio falls below $\sim$0.9, though, the
increased amount of energy and angular momentum addition
that would be needed to spin up material into a Keplerian
disk is large enough to greatly restrict the number and type
of potential sources.
Townsend et al.\  (2004) suggested that gravity darkening
effects could be strong enough to make a distribution of
nearly critical rotation speeds {\em appear} to be shifted down
to values of 0.5--$0.8 V_{\rm crit}$ if the line profiles
were interpreted as if the stars were spherical.
The inclusion of gravity darkening, however, tends to complicate
the analysis to the extent that a unique determination of
$V_{\rm eq} \sin i$ from a single measured line width (for
any individual star) does not seem to be possible.

This paper attempts to disentangle the above effects by
using Monte Carlo forward modeling to produce a large
number of trial probability distributions of $V_{\rm eq}$.
Each distribution is processed, assuming random inclination
angles (and with inclination-dependent line narrowing due to
gravity darkening), to simulate an observed statistical
sample of line widths.
The most likely intrinsic distribution of Be-star rotation
speeds is thus determined by searching for the models with
the minimum $\chi^{2}$ differences between the simulated and
observed line width distributions.
The derived distributions of $V_{\rm eq}$, as a function of
spectral type, yield important empirical constraints on the
{\em threshold rotation speeds} for the occurrence of the
Be phenomenon.
This forward-modeling method is less ambiguous than the
more common inverse technique of using simple geometric
transformations to convert an observed distribution
of $V_{\rm eq} \sin i$ values into either a distribution of
intrinsic rotation speeds or a mean value for $V_{\rm eq}$.

Although this kind of analysis has a long history (e.g.,
Chandrasekhar \& M\"{u}nch 1950; Stoeckley 1968; Lucy 1974;
Balona 1975; Porter 1996; Clark \& Steele 2000;
Chauville et al.\  2001),
the present work contains several novel features that help
to increase the overall level of confidence in the results.
First, the number of observed stars---from the published
database of Yudin (2001)---is now large enough to be able
to use the detailed {\em shapes} of the number distributions
as constraints rather than just their low-order moments.
Second, the effects of gravity darkening are included in the
most ``conservative'' manner possible, thus taking into account
the heterogeneous origins of the $V_{\rm eq} \sin i$ entries
in the database (i.e., gravity darkening was considered in the
calculation of some $V_{\rm eq} \sin i$ values, but not others).
The resulting subcritical values of $V_{\rm eq}$ are thus
designed to be safe upper limits, and the actual rotation speeds
may be even {\em lower} if the modeled gravity darkening effects
were overestimated.
Third, the derived rotation speeds are used as inputs to an
independent statistical simulation of visible polarization
measurements of Be-star disks.
The good agreement between the shapes of the observed and
simulated distributions of polarization is a useful validation
of the derived range of $V_{\rm eq}$ values.

The remainder of this paper is organized as follows.
{\S}~2 presents a summary of the Yudin (2001) Be-star database
and a description of the adopted fundamental stellar parameters
that were used to compute $V_{\rm crit}$ and other physical
quantities.
{\S}~3 describes the Monte Carlo forward modeling procedure
that was used to simulate statistical distributions of Be stars,
and also gives the resulting best-fit ranges of equatorial
rotation speed.
The possibility that nonstandard gravity darkening exponents
may apply to Be stars is investigated in {\S}~4, and a
simulation of linear polarization values for the derived
distribution of rotation speeds is presented in {\S}~5.
Additional pieces of evidence in favor of the results 
derived in {\S}~3 (essentially that
$V_{\rm eq} \neq V_{\rm crit}$ for early-type Be stars)
are laid out in {\S}~6.
A summary of the major results of this paper, together with
a discussion of the implications for theories of the Be
phenomenon, is given in {\S}~7.

\section{The Observational Database}

The Yudin (2001) database of early-type emission-line stars
contains 627 objects with MK spectral types between O7.5e and
B9/A0e and luminosity classes between II/III and V.
Care was taken to exclude Of, B[e], and Herbig Ae/Be stars from
this database, thus making it the largest sample of ``classical''
Be stars yet assembled.
There are 462 stars in the catalog\footnote{%
Note that Yudin (2001) states that there are 463 stars that have
nonzero values of the projected rotation speed, but the online
version of the database (VizieR catalog J/A$+$A/368/912) appears
to contain only 462 nonzero values of $v \sin i$.  This
discrepancy is unimportant for any of the statistical results of
Yudin (2001) or this paper.}
with nonzero values for the projected rotation speed $v \sin i$,
and this subsample contains the primary observational data to
be compared with the Monte Carlo model predictions in {\S}~3.
(The lower-case $v$ is used here only in combination with
$\sin i$ to denote the convolved quantity derived empirically
from line widths.
The nomenclature $V_{\rm eq} \sin i$ is used below only for
the product of two known quantities.)

In order to remove potential biases arising from the substantial
variation of stellar parameters from the late-O to early-A
spectral ranges, the observed values of $v \sin i$ for each star
should be normalized by the star's critical rotation speed.
Fundamental parameters (e.g., mass $M_{\ast}$ and polar radius
$R_p$) for each star are needed to compute the critical
rotation speed, which is defined for a rigidly rotating
Roche-model star as
\begin{equation}
  V_{\rm crit} \, \equiv \,
  \sqrt{\frac{2 G M_{\ast}}{3 R_p}} \,\, ,
  \label{eq:Vcrit}
\end{equation}
with $G$ being the Newtonian gravitation constant (see, e.g.,
Jeans 1928; Collins 1963; Tassoul 1978).
The above expression is consistent with the existence of
continuum radiation pressure as long as von Zeipel (1924)
gravity darkening applies and the continuum Eddington factor
$\Gamma$ is less than $\sim$0.5 (Glatzel 1998;
Maeder \& Meynet 2000a).
For completeness, the Eddington factor is given by
\begin{equation}
  \Gamma \, = \, \frac{\sigma_{e} L_{\ast}}{4\pi c GM_{\ast}}
  \, \approx \, 2.7 \times 10^{-5} \,
  \frac{(L_{\ast}/L_{\odot})}{(M_{\ast}/M_{\odot})} \,\, ,
\end{equation}
where $c$ is the speed of light in vacuum, $\sigma_e$ is the
Thomson scattering opacity, and $M_{\odot}$ and $L_{\odot}$
are the Sun's mass and bolometric luminosity.
The numerical factor above was computed from a standard
solar abundance mixture ($X = 0.73$, $Y = 0.24$).
For B-type stars, $\Gamma$ is typically much smaller than 1,
thus justifying the choice of solution branch to the radial
force balance equation (at the surface of the critically
rotating star) that is implied by equation~(\ref{eq:Vcrit}).
The largest value of $\Gamma$ in Yudin's (2001) database
is 0.23 (for the O7.5 III star 68 Cyg), and there are only
3 other stars out of 627 that have $\Gamma \geq 0.1$.

The spectral types and luminosity classes listed in Yudin's (2001)
catalog were used to compute bolometric luminosities $L_{\ast}$
and mean effective temperatures $T_{\rm eff}$ by using the
statistical relations of de Jager \& Nieuwenhuijzen (1987).
Their mean tabulated values were derived from a set of 199
high-precision determinations of $L_{\ast}$ and 268
high-precision determinations of $T_{\rm eff}$ across the
Hertzsprung-Russell (H-R) diagram.
For uncertain spectral types and luminosity classes that were
listed by Yudin (2001) using two possible values (e.g.,
``B9/A0'' or ``III/IV''), the luminosities and temperatures
were computed for each value then averaged logarithmically.
Stars without a listed luminosity class were assumed to be
main sequence (class V) objects.
Although more recent calibrations of $L_{\ast}$ and
$T_{\rm eff}$ exist for the earliest-type stars (see, e.g.,
Garcia \& Bianchi 2004; Crowther 2005), the
de Jager \& Nieuwenhuijzen (1987) tables remain the most
complete and cohesive set of correspondences that is
separated by luminosity class.

Once $L_{\ast}$ and $T_{\rm eff}$ were determined for each
star, the stellar mass $M_{\ast}$ was computed by interpolating
between the evolutionary tracks published by Claret (2004).
First, the abscissa in the H-R Diagram was transformed from
$T_{\rm eff}$ to the scaled variable
$\Xi = T_{\rm eff}/L_{\ast}^{0.16}$, which skews the
main sequence to be roughly vertical.
Each of the 30 evolutionary tracks (spanning initial
masses between 0.8 and 126 $M_{\odot}$) was searched for
the point where $\Xi$ matched that of the star in question.
The luminosities and masses at these points were saved
into one-dimensional arrays, thus effectively giving a
tabulation of $M_{\ast}$ as a function of $L_{\ast}$.
The actual stellar mass was then computed by linear
interpolation using the empirically determined value
of $L_{\ast}$.
The resulting mass-luminosity relationships for luminosity
classes III, IV, and V were fit with simple quadratic
functions (in logarithm space) and are given here for
comparison to other calibrations:
\begin{equation}
  y = 0.2084 + 0.0634 x + 0.0325 x^{2}
  \,\,\,\,\,\, (\mbox{III})
\end{equation}
\begin{equation}
  y = 0.2055 + 0.0746 x + 0.0322 x^{2}
  \,\,\,\,\,\, (\mbox{IV})
\end{equation}
\begin{equation}
  y = 0.1997 + 0.0844 x + 0.0312 x^{2}
  \,\,\,\,\,\, (\mbox{V})
\end{equation}
where
$y = \log_{10} (M_{\ast} / M_{\odot})$ and
$x = \log_{10} (L_{\ast} / L_{\odot})$.
These fits apply only to the late-O to late-B range represented
in Yudin's (2001) database (i.e., masses between 2.5 and 38
$M_{\odot}$) and the fits are accurate to within 5\% in the
mass over this range.

Figure 1 shows the spectral type dependence of $V_{\rm crit}$
as computed from equation (\ref{eq:Vcrit}) for luminosity
classes III, IV, and V.
Corresponding values of the critical rotation speed given
by Porter (1996) and Yudin (2001) are also shown.
For main sequence stars, the values computed for this paper
are in good agreement both with these other plotted values and
with the often-cited tables of Schmidt-Kaler (1982),
Underhill (1982), Harmanec (1988), and Andersen (1991).
For luminosity classes III and IV, the computed $V_{\rm crit}$
values are systematically larger than those of Yudin (2001).
This is the result of the trend that the luminosities of
de Jager \& Nieuwenhuijzen (1987) tend to be on the low side
when compared with other calibrations for giants and subgiants.
For constant $T_{\rm eff}$, the stellar radius computed with a
lower luminosity would be smaller, and thus $V_{\rm crit}$
would be larger.

Figure 1 also displays the $V_{\rm crit}$ values computed by
Chauville et al.\  (2001) for each of the 116 Be stars in their
published database.
These values are plotted with the same symbols for all luminosity
classes, and it can be seen clearly that they are about 20\%
lower than the values derived in this work and on average they
are even slightly lower than Yudin's (2001) values.
The reasons for this systematic discrepancy are not clear, although
it seems to be at the root of the significant difference between 
the derived values of the mean ratio $v \sin i / V_{\rm crit}$
of Yudin (2001) and Chauville et al.\  (2001); see below.

There are two main pieces of evidence that support the comparatively
large values for $V_{\rm crit}$ derived in this paper:
\begin{enumerate}
\item
Recent model atmosphere based determinations of B-star
fundamental parameters (Fitzpatrick \& Massa 2005) give critical
rotation speeds that agree well with the curves shown in
Figure 1.
For the 5 luminosity class III stars (ranging from B3 to A0) in
the Fitzpatrick \& Massa (2005) database, the computed values of
$V_{\rm crit}$ were all slightly {\em larger} than the values
given in Figure 1.
For the 20 (17) stars of class IV (V), the computed $V_{\rm crit}$
values fell roughly equally above and below the respective curves
in Figure 1.
\item
When modeling the properties of Be stars above the main sequence
(i.e., luminosity classes III and IV), it is probably safest to
choose the set of critical rotation speeds that cleaves
{\em closest} to the main sequence values.
It has been known for some time that rapid rotation and gravity
darkening can make a main sequence star appear to be up to a
full magnitude brighter than than its surface-averaged luminosity
would indicate (e.g., Maeder \& Peytremann 1970;
Collins \& Sonneborn 1977; Collins et al.\  1991).
Despite some evidence that the Be phenomenon occurs
only during relatively late evolutionary phases, it is possible
that the true distribution of Be stars on the H-R diagram
remains closer to the main sequence than is implied by the
inferred fractions of luminosity class III and IV stars
(see also Fabregat \& Torrej\'{o}n 2000).
Therefore, it may err on the side of caution to use the present
calibration, which yields relatively high values of $V_{\rm crit}$
for Be stars classified as III and IV.
\end{enumerate}
More work needs to be done to more accurately pin down the
fundamental parameters of Be stars.
By default, the statistical models presented below mainly use the
high-end $V_{\rm crit}$ values derived for this paper
(sometimes denoted as ``Cranmer''), but a parallel analysis is
also done using the low-end Chauville et al.\  (2001) values.

The 462 stars with measured projected rotation speeds were
normalized by dividing Yudin's (2001) tabulated $v \sin i$ by
the computed values of $V_{\rm crit}$ for each star.
The mean value of the ratio $v \sin i / V_{\rm crit}$ was
found to be 0.485, which is slightly smaller than the value
of 0.50 found by Yudin (2001) using lower critical rotation
speeds.
Also, the latter value was computed as the ratio of the mean
$v \sin i$ to the mean $V_{\rm crit}$, not as the mean value
of the set of ratios.
The dimensionless standard deviation, skewness, and kurtosis
of the distribution of ratios (as defined by Press et al.\  1992)
were found to be 0.167, $-0.00106$, and $-0.432$, respectively.
The significantly negative kurtosis denotes a flat-topped
plateau-like distribution that seems to occur because the full
sample of stars is made up of subpopulations with a range of
mean values that sweep across the central peak of the full
distribution (see below).
This result highlights the usefulness of using the ratio
$v \sin i / V_{\rm crit}$ rather than just $v \sin i$ itself,
because Yudin (2001) found that the observed distribution of
$v \sin i$ values was well-represented by a normal distribution.
Thus the ``normality'' of the $v \sin i$ distribution could
be related more to the large spread in $V_{\rm crit}$ across
the B-type spectral range than to the true distribution of
relative rotation rates.

To facilitate the study of the onset of the Be phenomenon as
a function of spectral type, the 462 stars in the database
were divided into five subpopulations, two containing 93 stars
and three containing 92 stars.
Table 1 contains a summary of the properties of these subsamples,
which were generated by binning stars of similar $T_{\rm eff}$.
From hot to cool effective temperatures, the groups are labeled
very early (VE), early (E), medium (M), late (L), and
very late (VL).
The exact values of $T_{\rm eff}$ used for the dividing lines
between the groups were chosen in order to give all five
subpopulations a roughly equal number of stars.
Table 1 also lists the mean, standard deviation, skewness,
and kurtosis of the $v \sin i / V_{\rm crit}$ distributions
for each subpopulation.
For further statistical analysis as a function of spectral
type and luminosity class, see Yudin (2001).

Figure 2 displays the number distribution of $v\sin i /V_{\rm crit}$
for all 462 Be stars, as well as the distributions for the five
subpopulations defined in Table 1, each normalized to an integral
of unity (i.e., expressed as binned probability densities).
As noted by Yudin (2001), the mean values of
$v \sin i / V_{\rm crit}$ have a systematic dependence on
$T_{\rm eff}$, increasing monotonically from $\sim$0.4 for the
hottest (VE) stars to $\sim$0.6 for the coolest (VL) stars.
There is no clear spectral type dependence on the standard
deviations (or higher moments) of the subpopulation
distributions.
The distributions are plotted by dividing the region between
$v \sin i / V_{\rm crit} = 0$ and 1 into 11 equally spaced bins;
this provides a good balance between resolution and statistics.
Note that the smallest and largest computed values of
$v \sin i / V_{\rm crit}$ for the full database are 0.035
and 0.912.
(The fact that there are {\em no} stars with measured $v \sin i$
that exceed $V_{\rm crit}$ tends to support the conclusion
reached below that Be stars are not all rotating critically;
see also {\S}~4).

As an alternate derivation of the normalized projected rotation
speeds, the 462 $v \sin i$ values were also divided by the
{\em lower} $V_{\rm crit}$ values indicated by
Chauville et al.\  (2001).
A least-squares power-law fit was found for the spectral type
dependence of the critical rotation speeds from that paper:
$V_{\rm crit} = (450 \,\, \mbox{km} \,\, \mbox{s}^{-1})/s^{0.775}$,
where $s$ is a continuous variable that straightforwardly denotes
the spectral subtype (for O5, B0, and B9, $s = 0.5$, 1.0, and 1.9,
respectively).
This fit does not reflect the spread in values from the range
of luminosity classes (see Figure 1) but it accurately models
the trend for the Chauville et al.\  critical speeds to be
lower than those derived by others.
Table 1 gives the mean, standard deviation, skewness,
and kurtosis of the $v \sin i / V_{\rm crit}$ distributions
for each subpopulation using this alternate calibration for
$V_{\rm crit}$.
For the earliest spectral bin, VE, the mean ratio of 0.49 is only
slightly higher than the value of 0.41 that corresponds to the
$V_{\rm crit}$ calibration derived above.
The differences between the two calibrations grow steadily larger
for the cooler spectral bins, leading to mean ratios of either
0.90 (Chauville et al.) or 0.59 (Cranmer) for the VL subpopulation,
depending on $V_{\rm crit}$.
For the entire database, the mean ratio using the Chauville
et al.\  calibration for $V_{\rm crit}$ is 0.627; this is
very close to the value of 0.65 given by
Chauville et al.\  (2001) for their sample of 116 Be stars.
This value is almost 30\% higher than the mean ratio of 0.485 that
was computed above, and the difference is due solely to the
differences in stellar mass and radius that go into the
calculation of the critical rotation speed.

\section{Forward Modeling of Distributions of the Projected
Rotation Speed}

In this section the procedure for simulating theoretical
distributions of $v \sin i / V_{\rm crit}$ for a random sample
of Be stars is described.
The simulation of $v \sin i$ measurements for individual
gravity darkened stars is outlined in {\S}~3.1 (see also
Collins 1974; Collins \& Truax 1995; Townsend et al.\  2004).
The construction of theoretical probability distributions for
a large number of stars, together with the comparison with the
observed distributions, is described in {\S}~3.2.

\subsection{Modeling Individual Stars}

For a given set of fundamental stellar parameters (mass, luminosity,
and polar radius), an equatorial rotation rate ($V_{\rm eq}$),
and an observational inclination angle ($i$), it is possible to
compute unique absorption line profile shapes that can be used
as direct measures of the quantity $v \sin i / V_{\rm crit}$.
Figure 3 illustrates the effect of oblateness and
von Zeipel (1924) gravity darkening on the absorption line
shape for a representative late-B main sequence star (see
also Cohen et al.\  2005).
As expected, gravity darkening preferentially weights the more
slowly rotating polar regions and thus produces a narrower
line profile than that computed without gravity darkening.
To produce the images and profiles in Figure 3, the computer
code from Cranmer \& Owocki (1995) was used to generate
high-resolution models of both a spherical star and an oblate
Roche-model star with the same polar radius and equatorial
rotation velocity.
The stars are observed from a large distance (1000 $R_p$) at a
given inclination angle, and the same limb darkening law is
applied in both cases (see below).
Each pixel of the image has a projected linear dimension of
0.008 $R_p$ in the plane perpendicular to the line of sight that
crosses through the center of the star.

The rotationally broadened profiles, here assumed to be the
commonly used \ion{He}{1} $\lambda$4471 absorption lines,
were simulated for the two cases shown in Figure 3 by the
following procedure.
The residual flux is defined as the ratio of the flux
in the spectral line to that of the surrounding continuum,
\begin{equation}
  {\cal R}_{\nu} \, = \,
  \frac{\int dx \, dy \, I_{\nu, {\rm L}} (x,y)}
       {\int dx \, dy \, I_{\nu, {\rm C}} (x,y)} \,\, ,
  \label{eq:Rnu}
\end{equation}
where $x$ and $y$ are spatial coordinates that specify a grid of
rays parallel to the observer's line of sight.
For the subset of these rays that intercept the stellar surface,
Cranmer \& Owocki (1995) and Cranmer (1996) described how to
convert these coordinates into star-centered spherical coordinates
$r$ and $\theta$ (ignoring the azimuthal longitude $\phi$ because
the stars are assumed to be axially symmetric).
The colatitude $\theta$ determines the local magnitude of the
centrifugally modified effective gravity, and also---for the
oblate gravity-darkened case---the local effective temperature.

For simplicity, the specific intensity of the continuum is
modeled as the product of a Planck function at the local
effective temperature and a linear limb darkening function,
i.e.,
\begin{equation}
  I_{\nu, {\rm C}} \, = \, \frac{5}{4\pi} B_{\nu}(T_{\rm eff})
  \, (1 - u_{c} + u_{c} \mu)  \,\, ,
\end{equation}
where $u_{c}$ is the continuum limb darkening constant and
$\mu$ is the cosine of the angle between the line of sight and
the normal to the stellar surface.
Below, we adopt $u_{c} = 0.4$ from both a classical treatment
of limb darkening for a gray temperature distribution at a
continuum wavelength of 4471 {\AA} (see eq.~10-20 of Mihalas 1978)
and a numerical study of limb darkening for rapidly rotating B
stars (Table 4 of Collins \& Truax 1995).
The specific intensity in the spectral line $I_{\nu, {\rm L}}$
is modeled as the product of $I_{\nu, {\rm C}}$ and a residual
intensity $f_{\nu}$ that is computed from Milne-Eddington theory
for a pure absorption line, with
\begin{equation}
  f_{\nu} \, = \, \frac{1 - u_{c} + u_{c} \mu / (1+\eta_{\nu})}
  {1 - u_{c} + u_{c} \mu}
\end{equation}
(Mihalas 1978; Collins 1989).
The dimensionless line absorption coefficient $\eta_{\nu}$ is
modeled as a Voigt profile that is thermally broadened and
Doppler shifted by the stellar rotation,
\begin{equation}
  \eta_{\nu} \, = \, \eta_{0} \, H (a, \xi)  \,\, ,
\end{equation}
and
\begin{equation}
  \xi \, = \, \frac{\nu - \nu_{0} (1 + V_{\rm LOS}/c)}
  {V_{\rm th} \nu_{0}/c}
\end{equation}
where $H(a,\xi)$ is the standard Voigt function,
$\nu_0$ is the rest-frame frequency of the line transition,
$V_{\rm LOS}$ is the projected component of the rotation 
velocity along the line of sight for this ray, and $V_{\rm th}$
is the thermal speed of helium atoms corresponding to the local
effective temperature.
The dimensionless strength of the Voigt wings is assumed here to
be given by a constant value of $a = 0.08$ based on standard
Stark broadening for the \ion{He}{1} $\lambda$4471 line at a
mean B-star temperature of $2 \times 10^4$ K (e.g.,
Griem et al.\  1962; Barnard et al.\  1969; Leckrone 1971).

The strength of the absorption line for each ray that intercepts
the stellar surface is parameterized by the dimensionless
line-center absorption coefficient $\eta_{0}$.
This quantity is determined empirically from a grid of detailed
spectral synthesis calculations of the \ion{He}{1} $\lambda$4471
equivalent width from a collection of early-type atmosphere models
(Gonz\'{a}lez Delgado \& Leitherer 1999).
The effective gravity and temperature at the stellar surface for
each ray is used to interpolate the local equivalent width from
Table 6 of Gonz\'{a}lez Delgado \& Leitherer (1999).
The conversion between $\eta_0$ and equivalent width $W$ is given
by the theoretical curve of growth,
\begin{equation}
  W \, = \, \frac{2 u_{c} \nu_{0} V_{\rm th}}
  {c [(1-u_{c})\sqrt{3} + u_{c}]} \int_{0}^{\infty} d\xi \,
  \left[ \frac{\eta_{0} H(a,\xi)}{1+\eta_{0} H(a,\xi)} \right]
  \label{eq:CoG}
\end{equation}
where $W$ above is in frequency units (Mihalas 1978).
This relation has been computed numerically and tabulated
for the adopted values of $a$ and $u_c$.
For B stars, the equivalent widths for individual rays span
the range between 0.1 and 1.5 {\AA}, corresponding roughly to
$\eta_0$ values between 30 and 3000.

The procedure outlined in equations~(\ref{eq:Rnu})--(\ref{eq:CoG})
is certainly more simplistic and approximate than performing full
model atmosphere calculations of the relevant line profiles
(as was done by, e.g., Townsend et al.\  2004).
However, the relative ease of computing reasonably accurate profile
shapes by the above method allows {\em very fine grids} of
wavelengths, inclination angles, and rotation speeds to be
computed without prohibitive computational expense.
These grids make possible the detailed statistical studies
described below.

Spectral lines were computed for representative main sequence stars
of spectral types B0, B2, B3, B5, and B9; i.e., for stars roughly
at the centers of the five subpopulation bins defined in {\S}~2.
For each spectral type, line profiles for the spherical and the
oblate gravity darkened cases were computed on a fine grid of
200 wavelength points, and for 100 inclinations ($i$ between
{0\arcdeg} and {90\arcdeg}) and 100 equatorial rotation speeds
($V_{\rm eq}/V_{\rm crit}$ between 0 and 1).
A single measure of the line width was determined from the full
width at half maximum (FWHM) of the numerically computed
profiles.
The key parameter to be used below is the dimensionless ratio $D$,
which is defined as the FWHM line width for a model star computed
with oblateness and gravity darkening divided by the corresponding
FWHM for the spherical model.
Thus, $D$ provides an indication of the relative change in the
absorption line width that is produced by the effect of
gravity darkening.
Figure 4 shows $D$ as a function of both $V_{\rm eq}/V_{\rm crit}$
and $i$ for the earliest and latest (B0 and B9) spectral types.
The information contained in Figure 4 is essentially equivalent
to that shown in, e.g., Figure 4 of Stoeckley (1968), Figure 6 of
Collins \& Truax (1995), Figure 1 of Townsend et al.\  (2004),
and Figure 7 of Fr\'{e}mat et al.\  (2005), but for different
spectral types.
These other figures have tended to plot the computed line widths
as functions of {\em projected} rotation speeds, which highlights
the ambiguity involved with attempting to determine the product
$V_{\rm eq} \sin i$ from from an ``observed'' $v \sin i$.
The forward modeling procedure outlined in this paper is
essentially free of such ambiguity.

The overall impact of the line narrowing due to gravity darkening
can be understood better by converting from $D$ to the similarly
defined ``velocity deficiency'' quantity of
Townsend et al.\  (2004),
\begin{equation}
  \delta V \, = \, (1-D) V_{\rm eq} \sin i / V_{\rm crit}
  \,\, ,
\end{equation}
specifically for their fiducial case of edge-on inclination
($i={90\arcdeg}$) and $V_{\rm eq}/V_{\rm crit} = 0.95$.
For the five spectral subtypes, there is a gradual increase of
the effect as one goes from B0 ($\delta V = 12.5$\%) to
B9 ($\delta V = 26.0$\%), with intermediate values of 16.0\%,
17.5\%, and 21.0\% for B2, B3, and B5, respectively.
These values compare favorably with those given in Table 2 of
Townsend et al.\  (2004), which helps to validate the use of
the simpler line synthesis technique described above.

There are noticeable differences between the shapes of the
contours in the B0 and B9 cases shown in Figure 4.
The intermediate (B2, B3, B5) spectral types have contours in
$D$ that change in shape gradually between the two plotted
extreme cases.
These differences arise because the \ion{He}{1} $\lambda$4471
equivalent widths, as interpolated from the modeled spectra
of Gonz\'{a}lez Delgado \& Leitherer (1999), depend on latitude
differently for different ranges of effective gravity and
temperature.
An early (B0) model rotating at 99\% of critical exhibits a maximum
in $W$ at mid-latitudes of about 0.9 {\AA}, and minima at the
poles and equator of $\sim$0.75 and $\sim$0.3 {\AA}, respectively.
A similar late-type (B9) model exhibits a simpler monotonic decrease
in $W$ from pole ($\sim$0.35 {\AA}) to equator ($\sim$0.1 {\AA}).
These differences result in different spectral line shapes and
different intensity weightings over the oblate surfaces. 

The procedure to simulate a measurement of $v \sin i/V_{\rm crit}$
for an individual star is summarized by the following relation:
\begin{equation}
  \left( \frac{v \sin i}{V_{\rm crit}} \right)_{\rm \! meas}
  \, = \, \left( \frac{V_{\rm eq}}{V_{\rm crit}} \right)
  D \sin i \, (1 + \zeta)  \,\, .
  \label{eq:vsinimeas}
\end{equation}
The choice of a range of values for $V_{\rm eq}/V_{\rm crit}$ is
discussed below in {\S}~3.2, and the inclination angle $i$ is
defined formally as $\cos^{-1} \mu_i$, where $\mu_i$ is chosen
randomly from a uniform probability distribution between 0 and 1
(see also Chandrasekhar \& M\"{u}nch 1950).
The gravity darkening factor $D$ is interpolated from the grids of
values that were used to generate Figure 4.
The factor $\zeta$ above is a simulated observational uncertainty,
which is sampled from a normal random distribution having a mean
of zero and a standard deviation of $\sigma_{\zeta}$
(i.e., 68\% of the time, $\zeta$ falls between $-\sigma_{\zeta}$
and $+\sigma_{\zeta}$).
The ``mean uncertainty level'' $\sigma_{\zeta}$ is kept constant
for any given sample of model stars, though the effects of
varying this parameter between 0 and 0.3 are explored below
(see also Balona 1975).
Yudin (2001) discussed the determination of standardized error
bounds for the 462 observed $v \sin i$ values in the database,
and found typical relative uncertainties of order 10\%.
Uncertainty levels up to 30\% in the ratio $v \sin i/V_{\rm crit}$
may be reasonable to assume, since neither the numerator nor the
denominator are known precisely.

In order to most stringently test the hypothesis that Be stars are
rotating nearly critically, it would be desirable to make the
assumptions that are most favorable to that hypothesis (i.e.,
those that tend to give largest derived values of
$V_{\rm eq}/V_{\rm crit}$).
Thus, if the derived rotation speeds are still substantially below
critical even when those assumptions are made, the hypothesis of
critical or nearly critical rotation can be ruled out at a
high level of confidence.
Two such assumptions regarding the use of $D$ in
equation~(\ref{eq:vsinimeas}) are adopted here:
\begin{enumerate}
\item
Despite the variations of $D$ as a function of spectral type,
in the statistical models below we apply the grid of $D$ values
computed for the B9 case to {\em all} of the spectral-type
subpopulations.
This is done because the B9 case shows the strongest line-narrowing
effect due to gravity darkening (i.e., the lowest values of $D$
for rapid rotation and edge-on inclination).
Applying this lower-limit case for $D$ in
equation~(\ref{eq:vsinimeas}) leads to a possible overestimate
of the narrowing trend and thus a systematic underestimate of
$v \sin i / V_{\rm crit}$ for a specific assumed $V_{\rm eq}$.
When compared with observed values of $v \sin i / V_{\rm crit}$,
then, the resulting {\em derived} $V_{\rm eq}$ would end up being
an overestimate.
\item
The fact that $D$ is included at all in the equation above implies
that the reported measurements of $v \sin i / V_{\rm crit}$ have
{\em not} taken gravity darkening into account.
However, several of the sources of observational $v \sin i$
data used by Yudin (2001) certainly did their best to account
for gravity darkening, and thus they reported processed estimates
of $V_{\rm eq} \sin i$ rather than raw measurements of $v \sin i$.
If this is the case for a substantial fraction of the observations,
then the use of equation~(\ref{eq:vsinimeas}) would lead to a
systematic underestimation of $v \sin i / V_{\rm crit}$.
This effect works in the same sense as the overestimation of the
strength of line narrowing from using the B9 models for all
spectral types.
\end{enumerate}
Thus, the present models stand at one end of a continuum of
modeling choices (i.e., ``strong'' gravity darkening effects),
and the assumption of {\em no} gravity darkening (i.e., $D = 1$)
would stand at the opposite end.
The true rotation speeds of the Be stars should fall somewhere
between the values derived for these limiting cases.

\subsection{Monte Carlo Distributions and Results}

There have been numerous attempts to deconvolve intrinsic
statistical distributions of stellar rotation speeds from the
observed distributions of $v \sin i$ values.
Analytic studies include Chandrasekhar \& M\"{u}nch (1950),
Bernacca (1970, 1972), Lucy (1974), and Balona (1975).
More recent efforts to determine the statistical distributions
numerically---rather than just estimate mean values or
assume that all stars have the same rotation rate---include
Wolff et al.\  (1982), Chen \& Huang (1987),
Porter (1996), and Brown \& Verschueren (1997).
Generally, these models tend to be ``inverse'' determinations
that begin with an observed distribution function of projected
rotation speeds $\Phi (v \sin i)$ and work iteratively towards
a consistent form for the true distribution function of
intrinsic rotation speeds $\Psi (V_{\rm eq})$.
This technique becomes potentially ambiguous, though, when
the line-narrowing effects of gravity darkening are taken
into account; i.e., how are we to be sure that the derived
solution for $\Psi (V_{\rm eq})$ is {\em unique} when an
observed $v \sin i$ cannot be mapped identically to a single
value of the product $V_{\rm eq} \sin i$?

Although the present forward-modeling method does not produce
completely unique solutions, it treats all possible distributions
on the same footing and thus is not in danger of missing
potential solutions.
After choosing a parameterized functional form for the distribution
of rotation speeds $\Psi(V_{\rm eq})$, the parameters are varied
by producing a multidimensional grid of trial distributions
having the full range of combinations of parameters.
(All velocities here are assumed to be in units of $V_{\rm crit}$.)
Each trial distribution is converted into an observed distribution
$\Phi (v \sin i)$ by using equation~(\ref{eq:vsinimeas}),
and each of these is compared to the observed distributions
for the five spectral-type subsamples shown in Figure 2.
The best fits are judged with the $\chi^2$ diagnostic
appropriate for comparing a coarsely sampled observed
distribution with a known model distribution (see below).

Earlier studies have used various functional forms for
$\Psi(V_{\rm eq})$ such as Dirac delta functions, Gaussians, and
uniform (i.e., flat-topped) distributions.
After some experimentation, it was found that a truncated linear
function of the form
\begin{equation}
  \Psi (V_{\rm eq}) \, = \, \left\{
  \begin{array}{ll}
   m V_{\rm eq} + b \,\, , &
   V_{\rm min} \leq V_{\rm eq} \leq V_{\rm max} \\
   0 \,\, , & \mbox{otherwise,} \\
  \end{array}
  \right.
  \label{eq:psidef}
\end{equation}
best balanced the demands of versatility and simplicity.
(From their appearance when plotted, these distributions can also
be called ``trapezoidal.'')
The three free parameters of this family of functions are
the minimum and maximum truncation values
($V_{\rm min}$ and $V_{\rm max}$) and the slope $m$.
For any choice of these parameters, the constant $b$ is
determined automatically by assuring that $\Psi (V_{\rm eq})$ is
normalized to unity upon integration over all $V_{\rm eq}$.
The minimum and maximum possible slopes are defined by the
distribution being ``right triangular;'' i.e., the minimum slope
occurs when $\Psi (V_{\rm min})$ is the global maximum and
$\Psi (V_{\rm max})=0$.
The maximum slope occurs when $\Psi (V_{\rm max})$ is the global
maximum and $\Psi (V_{\rm min})=0$.
Thus it is convenient to define a dimensionless parameter $S$
that ranges between $-$1 and $+$1, with the slope ranging from
its minimum to its maximum value over this range, and
\begin{equation}
  m \, = \, \frac{2 S}{(V_{\rm max} - V_{\rm min})^2}  \,\, ,
\end{equation}
\begin{equation}
  b \, = \, \frac{1}{V_{\rm max} - V_{\rm min}} \left[
  1 - S \left( \frac{V_{\rm max} + V_{\rm min}}
                    {V_{\rm max} - V_{\rm min}}
  \right) \right] \,\, .
\end{equation}
The parameter $S$ tends to give a better qualitative impression
of the shape of the function than does $m$.

For each choice of the above parameters, the resulting
distribution function $\Psi$ then needs to be converted into a
distribution function $\Phi$ for the observed values of 
$v \sin i / V_{\rm crit}$.
Rather than simulating large numbers of stars for each point
in the three-dimensional grid of parameter space
($V_{\rm min}$, $V_{\rm max}$, $S$), a series of Monte Carlo
simulations was performed for a one-dimensional grid of
delta-function distributions,
\begin{equation}
  \Psi_{j} (V_{\rm eq}) \, = \,
  \delta (V_{\rm eq} - V_{j})  \,\, ,
\end{equation}
with 80 equally spaced values of $V_{j}/V_{\rm crit}$ ranging
between 0 and 1.
These distributions then served as {\em basis functions} used to
build up any desired arbitrary statistical distribution.
Each Monte Carlo simulation used 10$^5$ model stars
with random inclinations $i$ and observational uncertainties
$\zeta$ as described in {\S}~3.1, and identical values
of $V_{\rm eq} = V_{j}$.
Equation~(\ref{eq:vsinimeas}) was used to simulate individual
values of $v \sin i / V_{\rm crit}$, and these values were summed
into the same 11 bins as were used in Figure 2 for comparison
with observations.
This process yielded a a one-dimensional grid of
$\Phi_{j} (v \sin i)$ functions corresponding to the
delta-function $\Psi_{j}$ grid.
Finally, then, the $\Phi$ function corresponding to a given
trapezoidal distribution $\Psi$ (eq.~[\ref{eq:psidef}]) was
computed by taking the linear combination 
\begin{equation}
  \Phi (v \sin i) \, \propto \, \sum_{j}
  \Phi_{j} (v \sin i) \, \Psi(V_{j})  \,\, ,
\end{equation}
and renormalizing to an integral of unity.
In other words, the trapezoidal distribution $\Psi$ is used
as a weighting function to determine what fraction of each 
of the 80 $\Phi_j$ functions are summed together to form the
complete distribution $\Phi$.

Before presenting the results of the above process, it is
instructive to show the $\Phi_{j} (v \sin i)$ distributions
that correspond to $V_{j} = V_{\rm crit}$.
These represent the predicted distributions of projected rotation
speed assuming that {\em all Be stars rotate critically.}
Figure 5 shows a series of these distributions that were
computed for different choices for the mean uncertainty level
$\sigma_{\zeta}$.
For $\sigma_{\zeta} = 0$, the maximum modeled value of
$v \sin i / V_{\rm crit}$ is $\sim$0.68, corresponding
essentially to the maximum value of $D \sin i$ along the right
edge of Figure 4b.
(If gravity darkening had not been taken into account, the
maximum modeled value of $v \sin i / V_{\rm crit}$ for the
$\sigma_{\zeta} = 0$ model would have been 1.)
Note the secondary peak around $v \sin i / V_{\rm crit} \approx
0.53$; this corresponds to the small local maximum in $D$ for
$i \approx {40\arcdeg}$ and $V_{\rm eq}/V_{\rm crit} = 1$.
Predictably, when $\sigma_{\zeta}$ is increased, the probability
distributions become smeared out into increasingly Gaussian shapes.

Figure 5 also shows the observed $\Phi$ distribution of the
earliest-type (VE) subpopulation of Be stars, normalized by
the low choice of $V_{\rm crit}$ from the fit to the
Chauville et al.\  (2001) values.
(The distribution is shifted to the right compared to the
one shown in Figure 2b.)
Thus, the observed distribution is plotted with assumptions that
push the peak to high values of $v \sin i / V_{\rm crit}$ and
the modeled distributions are plotted with the
``strong gravity darkening'' assumptions that push the peaks
to low values.
From the fact that there is still a 15--20\% mismatch between
the observed and modeled peaks, despite good-faith efforts to
push them together, it seems evident that the early-type Be stars
cannot be all rotating with $V_{\rm eq} = V_{\rm crit}$.
If the self-consistent grid of $D$ values for the B0 spectral
type (shown in Figure 4a) was used instead of the lower-valued B9
grid, the modeled curves would be shifted further to the right
in Figure 5 by about 18\% (see vertical line in upper-right),
thus making critical rotation even more of a mismatch.

The quantitative degree of agreement between a simulated distribution
$\Phi (v \sin i)$ (which is known accurately) and a coarsely
sampled observed distribution is determined by using the $\chi^2$
diagnostic defined in {\S}~14.3 of Press et al.\  (1992):
\begin{equation}
  \chi^{2} \, \equiv \, \sum_{i=1}^{N_b}
  \frac{(n_{\rm obs} - n_{\rm mod})^2}{n_{\rm mod}}
\end{equation}
where $N_{b} = 11$ bins, and where $n_{\rm obs}$ and $n_{\rm mod}$
are the observed and modeled number counts in each bin,
respectively.
For the purposes of computing $\chi^2$, the normalized distributions
are multiplied by constant factors of either 92 or 93 in order
for them to contain the same total number of stars as the observed
subsample distributions.
For the various models discussed below, Table 1 presents
normalized ratios $\chi^{2} / \nu$, where $\nu = N_{b}-1$ is the
effective number of degrees of freedom.
Each resulting value of $\chi^2$ is related to the probability
that the observed counts are drawn from the same ``known''
distribution from which the modeled counts are drawn.
Assuming normally distributed uncertainties, this probability
is given by
\begin{equation}
  Q (\chi^{2} | \nu) = \frac{1}{\Gamma (\nu / 2)}
  \int_{\chi^{2} / 2}^{\infty} e^{-t} t^{(\nu / 2)-1} dt
\end{equation}
where $\Gamma(x)$ is the complete Gamma function.
Naturally, when $\chi^{2} \ll \nu$ the above probability
approaches unity (i.e., the modeled distribution is a good
match to the observations), and when $\chi^{2} \gg \nu$ the
above probability is negligibly small.

To locate the optimum range of $V_{\rm eq}$ values for each
subsample, the three free parameters of the trapezoidal
probability distribution $\Psi(V_{\rm eq})$ were varied
until global minima in $\chi^{2}/ \nu$ were found.
This process was repeated 7 times, once for each of the
following assumed values of the mean uncertainty level:
$\sigma_{\zeta} = 0$, 0.05, 0.10, 0.15, 0.20, 0.25, and 0.30.
In each case, the global minima in $\chi^{2}/ \nu$ were
located by: (1) automatically searching through the full
three-dimensional parameter space, and (2) making contour plots
of $\chi^{2}/ \nu$ in several two-dimensional slices through
the parameter space to make sure that no minima were missed.
Table 1 lists the best-fit parameters and optimal values of
$\chi^{2} / \nu$, $Q (\chi^{2} | \nu)$, and $\sigma_{\zeta}$
for the two assumptions concerning $V_{\rm crit}$ discussed
in {\S}~2 (i.e., high-end and low-end limits).
In all cases the best fits were found with either
$\sigma_{\zeta} = 0.15$ or 0.20.
Lower values of $\sigma_{\zeta}$ tended to produce
unrealistically sharp number distributions $\Phi (v \sin i)$,
and higher values produced distributions that were too broad.
This seems to be an independent verification that the
observational uncertainties of the Yudin (2001) $v \sin i$
values are about 10--20\%.

As seen in Table 1, there were two cases where no acceptable
fits could be found to the observed number distributions:
the L and VL subpopulations normalized by the low-end
Chauville et al.\  (2001) critical rotation speeds.
For these cases the peaks in the observed $\Phi (v \sin i)$
distributions occurred at values of $v \sin i/V_{\rm crit}$
larger than than 0.8, but the modeled statistical
distributions---even for critical rotation---have peak values
of $v \sin i/V_{\rm crit}$ no greater than $\sim$0.7, as seen
in Figure 5.
If the Chauville et al.\  values of $V_{\rm crit}$ are correct,
these late-type Be stars seem to be consistent with {\em nearly
critical rotation} as well as possibly a weaker line-narrowing
effect due to gravity darkening than has been modeled here.

For ease of comparison with other analyses, Table 1 also gives
the weighted mean ratios of equatorial rotation speed to critical
rotation speed for each subpopulation, defined by
\begin{equation}
  \langle V_{\rm eq}/V_{\rm crit} \rangle \, \equiv \,
  \left[ \int dx \, \Psi(x) \, x \right] \left/
  \left[ \int dx \, \Psi(x) \right] \right.  \,\, ,
\end{equation}
where $x = V_{\rm eq}/V_{\rm crit}$ is used for brevity.
Taking into account that the five subpopulations have
roughly equal numbers of stars, the mean ratio for the entire
database of 462 stars is computed simply by averaging together
the five values for each subpopulation.
For the high-end (Cranmer) and low-end (Chauville et al.) choices
of $V_{\rm crit}$ values, the weighted mean values of
$\langle V_{\rm eq}/V_{\rm crit} \rangle$ are 0.684
and 0.854, respectively.
The L and VL subpopulations that had no acceptable fits
for $\Psi(V_{\rm eq})$ in the low-end $V_{\rm crit}$ case
were assumed to be completely critically rotating, so the
above weighted mean value of 0.854 is an upper limit.
Fr\'{e}mat et al.\  (2005) derived a most probable ratio of
$V_{\rm eq}/V_{\rm crit} \approx 0.75$ from an analysis of
the 116 Be stars published by Chauville et al.\  (2001) plus
14 others.
This value falls nearly halfway between the two mean ratios given
above that were derived under the assumptions of lower and
upper limiting cases for $V_{\rm crit}$.

Figure 6 presents a summary of the best-fit probability
distributions $\Psi(V_{\rm eq})$ for each spectral-type
subsample and for both the high-end and low-end calibrations
of $V_{\rm crit}$.
Figure 6 also compares the simulated and observed distributions
of projected rotation speed $\Phi (v \sin i)$.
Several features of these plots are noteworthy:
\begin{enumerate}
\item
There is a definite dependence of $V_{\rm min}$ on spectral type.
The hottest Be stars (subsamples VE and E) seem to have lower
bounds on their rotation speed distributions that extend down
to 40--50\% of critical.
This lower bound increases, as $T_{\rm eff}$ decreases,
to the point where the L and VL subsamples are consistent with
nearly critical rotation.
An extremely approximate fit to the dependence of $V_{\rm min}$
on the stellar effective temperature is given by
\begin{equation}
  V_{\rm min} / V_{\rm crit} \, \approx \, 
  0.28 \, \tanh \left[ 13.3 - (T_{\rm eff} / 1500 \, \mbox{K})
  \right] + 0.72
  \label{eq:Veqmin}
\end{equation}
(see also Figure 8 below).
\item
The derived values of $V_{\rm max}$ do not vary systematically
with spectral type.
Indeed, a decent broad-brush approximation would be to
assume that $V_{\rm max} \approx V_{\rm crit}$, and that
the rotation speeds of Be stars of a given spectral type
range from a $T_{\rm eff}$-dependent minimum value up
to the critical rotation speed.
\item
The derived shapes of $\Psi(V_{\rm eq})$ depend rather strongly
on the adopted mean uncertainty level $\sigma_{\zeta}$.
The higher the value of $\sigma_{\zeta}$, the narrower the
resulting best-fit probability distribution.
This is understandable because the simulated distributions
$\Phi (v \sin i)$ are modeled essentially as a convolution
between two distributions: the intrinsic distribution of
$V_{\rm eq}$ values and the normally distributed spread of
uncertainties ($1 + \zeta$).
If all of these distributions were Gaussian in shape, a
specified width of $\Phi$ could be obtained for an infinite
number of choices for $\Psi$ and $\zeta$, as long as the
root-mean-squared sum of their widths equaled the
desired width of $\Phi$.
Any future attempts to simulate these kinds of number
distributions must be sure to model the observational
uncertainties as accurately as possible.
\end{enumerate}

It is also worth noting that the choice of trapezoidal
parameterization for $\Psi (V_{\rm eq})$ (eq.~[\ref{eq:psidef}])
was somewhat limiting because even the ``best fitting''
choices of the parameters did not yield perfect fits to the
observed distributions $\Phi (v \sin i)$.
For example, the strongly nonmonotonic behavior in the observed
E subsample, for $v \sin i / V_{\rm crit} \approx 0.6$--0.7,
limited the best-fit values of the probability $Q (\chi^{2} | \nu)$
to be never larger than 50\%.
A more complicated functional form for $\Psi (V_{\rm eq})$ may
have been able to fit this feature better, but the introduction
of too many free parameters can lead to unphysically complex
distributions.
An attempt was made to vary the shape of $\Psi (V_{\rm eq})$
iteratively by using a ``simulated annealing'' algorithm---i.e.,
adopting randomized changes only if they resulted in lower values
of $\chi^2$---following the spirit, if not the exact method,
of Lucy (1974).
The resulting best-fit distributions always tended toward sums
of several (at least 3) sharply peaked subdistributions, with
large ranges of intervening $V_{\rm eq}$ having zero contribution.
These distributions were judged to be unphysical, and the
more smoothly varying trapezoidal distribution was determined
to be the best balance of realism and parameter flexibility.

To develop a complete understanding of what ranges of
$V_{\rm eq}$ can actually be {\em ruled out} by this analysis,
it is not enough to plot only the best-fit distributions.
Figure 7 provides a summary of the goodness-of-fit probabilities
$Q (\chi^{2} | \nu)$ that were obtained both from the unconstrained
parameter variation and also from other constrained searches of
parameter space that examined only stars rotating faster than
certain threshold values of $V_{\rm eq}$.
Only results for the VE and M spectral subranges are shown, since
the intermediate E subrange exhibits similar probability curves as
the VE case and the L and VL cases are consistent with nearly
critical rotation (and thus show no interesting behavior as the
threshold lower limits are varied).
These curves show how the probabilities for each subsample
decrease when successively higher (i.e., more limited) ranges
of $V_{\rm min}$ and $V_{\rm max}$ are allowed.
The probabilities given in Table 1 for unconstrained searches
of parameter space are shown in Figure 7 as the maximum values
at the left edges of each plot.
It is clear from Figure 7a that the VE number distribution
$\Phi (v \sin i)$ cannot be fit adequately with values of
$V_{\rm eq}$ that all exceed $\sim$0.8---no matter the
choice of $V_{\rm crit}$ calibration.
On the other hand, the necessity of subcritical rotation
speeds for the M number distribution (in Figure 7b) depends
sensitively on the adopted $V_{\rm crit}$ calibration.
For the high-end (Cranmer) values, no good fits can be obtained
when $V_{\rm eq}$ is constrained to be greater than 0.8.
For the low-end (Chauville et al.) values, though, critical
rotation has just a slightly lower probability than nearly
any degree of subcritical rotation and thus cannot be ruled out.

The above constraints put on $V_{\rm min}$ represent
determinations of the threshold values of the equatorial
rotation speed for the occurrence of the Be phenomenon.
Figure 8 gives a summary of the $T_{\rm eff}$ dependence of
the threshold rotation speeds from Table 1 and 
equation~(\ref{eq:Veqmin}), as well as an indication of how
$V_{\rm min}$ changes for non-optimal values of $\sigma_{\zeta}$
(see caption for selection criteria).
It should be emphasized that these values were all computed
by assuming a relatively strong line-narrowing effect from
von Zeipel gravity darkening.
This assumption of $D < 1$ for fast rotation tends to
transform a given distribution $\Psi (V_{\rm eq})$ into a
distribution $\Phi (v \sin i)$ which is shifted more toward
lower values of $v \sin i$ than would have occurred if $D=1$.
If, though, $D=1$ had been assumed in the above parameter
optimization, the resulting values of $V_{\rm min}$
would have all been {\em smaller} than the values given
in Table 1 and Figures 6 and 8.
Therefore, the values of $V_{\rm min}$ derived here seem to be
conservative upper bounds on the threshold for forming
Be-star disks.
If the gravity darkening effects were overestimated, then
the $V_{\rm min}$ values may be lower, but it is difficult
to imagine how they could be substantially higher.

\section{Departures from von Zeipel gravity darkening?}

The statistical results described in {\S}~3 depend somewhat
sensitively on the presumed form of gravity darkening used
for the rapidly rotating stars.
For stars with $T_{\rm eff} \gtrsim 8000$ K it is traditional
to assume that the mainly radiative energy transfer in
the subsurface layers maintains the ideal von Zeipel (1924)
linear scaling between emergent flux $F$
and effective gravity $g$ (i.e., $F \propto g$).
Cooler stars that exhibit subsurface convection are well-known
to show weaker gravity darkening, with $F \propto g^{4\beta}$
and the gravity darkening exponent $\beta \lesssim 0.1$
(see, e.g., Lucy 1967; Osaki 1970; Anderson \& Shu 1977;
Claret 1998, 2000, 2003).
Hot stars, though, may exhibit departures from the fully
radiative gravity darkening value of $\beta = 0.25$ if they
are differentially rotating (Smith \& Worley 1974), if
they are in close binary systems undergoing mass transfer
(Unno et al.\  1994), or if there is some kind of shear-driven
turbulence in their outermost layers that might help to
redistribute the flux (e.g., Smith 1970;
Smith \& Roxburgh 1977).
Some observational evidence exists from eclipsing binary
light curves that $\beta$ could be greater than 0.25 in some
cases (Kitamura \& Nakamura 1988; Nakamura \& Kitamura 1992;
Djura\v{s}evi\'{c} et al.\  2003).

A general relationship between the bolometric flux $F$
and the effective gravity $g$ (the latter defined as the
magnitude of the vector sum of gravity and the outward
centrifugal force) is expressible as
\begin{equation}
  F (\theta) = \sigma_{B} T_{\rm eff}^{4} (\theta) \, = \,
  \left( \frac{L_{\ast}}{\Sigma_{4 \beta}} \right) g^{4 \beta}
  \,\, ,
\end{equation}
where $\theta$ is the colatitude measured from the pole,
$\sigma_B$ is the Stefan-Boltzmann constant, and $\beta$
is the exponent defined in terms of effective temperature
(i.e., $T_{\rm eff} \propto g^{\beta}$).
Small changes in $L_{\ast}$ as a function of the rotation
rate of the star are neglected (see, however,
Fr\'{e}mat et al.\  2005).
The parameter $\Sigma_{4 \beta}$ is a normalizing constant
that is a function of the rotation rate, but is
not a function of $\theta$.
It is important to perform this normalization properly,
since a key part of the $v \sin i$ analysis involves the
comparison of stars modeled with and without gravity darkening.
(In other words, it is important to know which latitudes are
brighter, and which are dimmer, when compared with a star
modeled without gravity darkening.)
$\Sigma_{4 \beta}$ is the surface-weighted integral of
$g^{4 \beta}$ itself, and
\begin{equation}
  \Sigma_{4 \beta} \, \equiv \, \oint g^{4 \beta} dS
  \, = \, 2\pi \int_{0}^{\pi} \frac{g^{4 \beta}
  R_{\ast}^{2} \sin\theta \, d\theta}
  {\cos\delta} \,\, ,
\end{equation}
where $\delta$ is the angle between the local effective gravity
and the radius vector ($\cos \delta = -g_{r}/g$), and both $g$
and the stellar radius $R_{\ast}$ are functions of $\theta$
(see, e.g., Slettebak 1949; Collins 1963; Maeder 1999).
The two limiting cases of $\beta=0$ (no gravity darkening)
and $\beta = 0.25$ (standard von Zeipel gravity darkening) give
the surface area of the oblate star ($\Sigma_0$) and the
surface-weighted effective gravity ($\Sigma_1$), respectively.
For rigidly rotating Roche-model stars, Cranmer \& Owocki (1995)
presented a parameterized fit of $\Sigma_1$ as a function of
the stellar rotation rate, and Cranmer (1996) gave a
similar fit for $\Sigma_0$.
For this paper, this function was also computed for an intermediate
amount of gravity darkening between none and the standard amount
($\beta = 0.125$), and also for an extreme amount ($\beta = 0.4$)
that was indicated by recent modeling of the light curves of
early-type eclipsing binaries TT Aur and V Pup
(Djura\v{s}evi\'{c} et al.\  2003).

The line-narrowing factor $D$ was computed for the two new
choices for $\beta$ and was compared with the values plotted in
Figure 4 for the B9 spectral subtype.
Figure 9a shows how the decrease in $D$ with increasing
$V_{\rm eq}$ becomes stronger for larger values of $\beta$.
All curves are plotted for a constant inclination angle of
$i= 90\arcdeg$.
Figure 9b illustrates how the simulated distribution of projected
rotation speeds $\Phi_{j} (v \sin i)$ shifts toward lower values
for stronger gravity darkening exponents.
This plot shows the distributions computed for a delta-function
distribution of critical rotation speeds ($V_{j} = V_{\rm crit}$)
and thus is comparable to Figure 5.
The curves in Figure 9b were computed assuming a mean
uncertainty level of $\sigma_{\zeta} = 0.15$.
Note that for the case of no gravity darkening ($\beta = 0$)
there would be a significant number of observations
of $v \sin i$ values that exceed $V_{\rm crit}$.
The fact that no such observations exist in the Yudin (2001)
database implies that gravity darkening {\em does} need to be
taken into account when processing the existing measurements
of $v \sin i$.
Figure 9b also shows the observed number distribution for the
VL subsample of late-type Be stars, as normalized by the
larger set of critical rotation speeds derived in {\S}~2.
The curve corresponding to $\beta = 0.25$ seems to be the most
consistent with the observations, although values $\pm 0.05$
would also be reasonable.
The shape of the observed probability distribution---especially
at its upper end---may thus be a good diagnostic of the degree of
gravity darkening that is present in a population of nearly
critical rotators.

\section{Linear Polarization}

Until this point there has not been much discussion of the
dense circumstellar disks around Be stars that presumably
exist only when $V_{\rm eq} > V_{\rm min}$.
A primary source of information about these disks is the
measurement of linear polarization produced at broadband
visible wavelengths (Coyne \& McLean 1982; Bjorkman 2000b)
by Thomson scattering of free electrons in the flattened
envelope.
The Yudin (2001) database contains 335 entries with nonzero
values of both $v \sin i$ and the visible polarization
fraction $P_V$.
Several aspects of the observed ``triangular'' distribution of
$P_V$ versus $v \sin i$ were not easily explainable in
terms of earlier models of Thomson scattering in circumstellar
disks.
Thus, the goal of this section is to use the stellar rotation
properties derived above to produce a more accurate simulated
distribution of polarization values and compare with the
measured values.
However, this is not an attempt to produce rigorous models of
the physical properties of Be-star disks, but only a general
consistency test for the validity of the derived ranges of
rotation rate.

Polarization values for a simulated distribution of Be stars
were computed from a slightly modified form of the
single-scattering approximation formulae of McDavid (2001).
The required properties for each star were found by first
computing three random quantities:
(1) the inclination of the rotation axis, assumed to be identical
to the disk axis, using the same procedure as outlined earlier;
(2) the stellar spectral type, which was specified by a
nonrotating value of $T_{\rm eff}$ and was sampled from a
cumulative probability distribution that was tabulated from
the full Yudin (2001) database of 627 stars; and
(3) the rotation rate $V_{\rm eq}/V_{\rm crit}$, which was
sampled from a flat distribution ($S=0$) between $V_{\rm min}$
as given by equation~(\ref{eq:Veqmin}) and $V_{\rm max} =
V_{\rm crit}$.
The remaining basic stellar parameters ($R_p$, $M_{\ast}$, and
$L_{\ast}$) were interpolated from the relationships
derived in {\S}~2, assuming that all stars are on the main
sequence (luminosity class V).
The equatorial values of the stellar radius $R_{\rm eq}$
and effective temperature $T_{\rm eq}$ were computed assuming
Roche oblateness and ideal von Zeipel gravity darkening
($\beta = 0.25$).
The disk temperature ($T_{\rm disk}$) was assumed to be
constant and equal to 0.75 times $T_{\rm eq}$ (McDavid 2001).

The disk itself was modeled as occupying a spherical sector
around the equatorial plane with an opening half-angle $\alpha$.
The disk extends from its inner edge (assumed to be coincident
with the star's equatorial radius $R_{\rm eq}$) to infinity with
a power-law dependence of electron density with radius,
\begin{equation}
  N_{e} (r) \, = \, N_{e,0} \, ( r / R_{\rm eq} )^{-\eta}
  \,\, .
\end{equation}
Following McDavid (2001), the constant value of $\eta = 3.1$
is adopted for all stars.
In an attempt to account for the large variation of fundamental
parameters across the B-type spectral range, the inner disk
density $N_{e,0}$ was assumed to depend on $T_{\rm eff}$ as
\begin{equation}
  N_{e,0} \, \approx \, 7 \times 10^{11}
  \left( \frac{T_{\rm eff}}{10000 \, \mbox {K}} \right)^{2.2}
  \,\, \mbox{cm}^{-3}  \,\, .
  \label{eq:Ne0}
\end{equation}
This relation was derived from a linear fit (in $\log T_{\rm eff}$
versus $\log N_{e,0}$) to a total of 23 measured inner disk densities
for Be stars with spectral types ranging between B0.5 and B8
(Waters et al.\  1987; McDavid 2001).\footnote{%
For another discussion of this trend, see Slettebak et al.\  (1992).
Note, though, that van Kerkwijk et al.\  (1995) did not find
such a trend in Be-star disk properties as derived from infrared
observations.}
Figure 10 shows the $T_{\rm eff}$ dependence of these values and
compares the measurements to the above fit.
The estimated mass densities $\rho$ were computed under the
assumption of complete hydrogen ionization; thus
$\rho = N_{e} m_{\rm H}$ where $m_{\rm H}$ is the mass of a
hydrogen atom.

To gauge the approximate validity of the derived inner disk
densities, Figure 10 also plots two other densities that are
expected to bracket these values from above and below.
The photospheric mass densities are computed from the criterion
that, in the stellar photosphere, the Rosseland mean optical
depth should have a value of approximately one:
\begin{equation}
  \tau_{\rm R} \, \approx \, \kappa_{\rm R} \rho H \, = \, 1
\end{equation}
where $H$ is the photospheric scale height (proportional to
$T_{\rm eff}/g$) and $\kappa_{\rm R}$ is the Rosseland mean
opacity (in cm$^2$ g$^{-1}$) interpolated from the extensive
tabulation of Kurucz (1992).
The resulting photospheric densities were compared with densities
from detailed model atmospheres and were found to agree
to within about $\pm$20\% (R.\  L.\  Kurucz 2004, private
communication).
Figure 10 also shows an upper limit for the mass density at the
sonic point of the equatorial outflow (i.e., the radius at which
the radial flow speed equals the sound speed $c_s$), i.e.,
\begin{equation}
  \rho_{\rm sonic} \, = \, \frac{\dot{M}}{4\pi c_{s} r^2}
\end{equation}
where the mass loss rate $\dot{M}$ was computed from the stellar
wind parameterization of Vink et al.\  (2000).
The use of these values presumes that the disk mass fluxes are of
the same order of magnitude as the polar wind mass fluxes (see
{\S}~5.1 of Bjorkman 2000a).
The radius $r$ is set to $R_{\rm eq}$, though it should be noted
that in a slowly expanding viscous decretion disk, the sonic
radius may be much larger (e.g., Okazaki 2001) and thus the
sonic density may be smaller.
The sound speed $c_s$ is assumed to be constant inside the disk,
\begin{equation}
  c_{s} \, = \, \sqrt{ \gamma k T_{\rm disk}
  / \mu m_{\rm H}}
\end{equation}
where the adiabatic exponent $\gamma$ is 5/3, $k$ is
Boltzmann's constant, and $\mu$ is the mean molecular weight
of the gas (we assume $\mu = 0.6$).
The fact that the inner disk density falls between the
photospheric density and the sonic density indicates that the
base of the disk is most likely highly subsonic (i.e., close
to the stellar surface), but is also sitting several scale
heights {\em above} the photosphere.

The opening half-angle $\alpha$ of the equatorial envelope was
computed from the basic theory of Keplerian accretion disks,
with
\begin{equation}
  \alpha \, = \, f \tan^{-1} ( c_{s} / V_{\rm Kep} )
  \label{eq:akep}
\end{equation}
(see, e.g., Pringle 1981).
In the above equation, $V_{\rm Kep} = (GM/r)^{1/2}$ is the
Keplerian azimuthal velocity at distance $r$, here computed
at a fiducial distance of $2R_{\rm eq}$.
The dimensionless constant $f$ is an order-unity correction factor
that is adjusted to produce an overall level of polarization
similar to what is observed.
The {\em shape} of the resulting distribution of $P_V$ versus
$v \sin i$ does not depend strongly on the value chosen for
$f$, as long as the same value is used for all stars in the
simulated sample.
In the models presented below, $f = 3$, which yielded a
realistic range of $\alpha$ angles between {3\arcdeg} and
{10\arcdeg}.
The polarization from an axisymmetric disk is proportional to
$\sin\alpha \, \cos^{2}\alpha$, and for the small angles assumed
here this is approximately a linearly increasing function
of $\alpha$.

The above properties are the necessary inputs to the
single-scattering formulae of McDavid (2001).
The straightforward $\sin^{2} i$ inclination dependence of
that model, though, has been replaced by a more accurate
relation that takes account of stellar occultation by a
thin disk (i.e., eqs.~[17] and [19] of Fox \& Brown 1991).
The McDavid (2001) model was also simplified in two ways:
(1) the relatively weak non-LTE effects for hydrogen were
ignored, and (2) the polarization was computed only at the
center of the $V$ filter band (5500 {\AA}) instead of using
the full bandpass function.
The simulated values of $v \sin i/ V_{\rm crit}$ for each star
were computed by using equation~(\ref{eq:vsinimeas}) with
$\sigma_{\zeta} = 0.15$.

In Figure 11, the observed and simulated distributions of
$P_V$ are compared with one another for statistical samples
of 335 stars each.
The dotted triangular limits are shown only as a rough
envelope of the $v \sin i$ dependent spread of values, and
are the same in both plots.
Despite minor differences, the shapes of the observed and
modeled distributions are similar.
The rising upper envelope at low values of $v \sin i$
can be understood from the approximate $\sin^{2} i$
dependence of the polarization, since the lowest projected
rotation speeds correspond to the lowest inclination angles
(see also McLean \& Brown 1978).
The decrease in the upper envelope for the largest values of
$v \sin i$, however, is not so easy to interpret.
Yudin (2001) suggested three possible explanations for this
decrease in $P_V$ spread for the fastest rotators.
Two of these effects (depolarization from the oblate
gravity-darkened star itself, and lower intrinsic polarization
for disks around the later-type stars that also have the
highest values of $V_{\rm eq}$) were not thought to be strong
enough to produce the observed decreases.
The third proposed effect was that the opening angle $\alpha$
may be inversely proportional to $V_{\rm eq}$.
Whereas something like this may have been true for the
wind-compressed disk model of Bjorkman \& Cassinelli (1993),
the thickness of a true viscous decretion disk seems most
likely to become relatively uncoupled from the rotational
properties of the central object (see eq.~[\ref{eq:akep}]).

The simulated distribution of $P_V$ values shown in Figure 11b
does seem to match the observed strong decrease in spread at
the largest values of $v \sin i$ in Figure 11a.
This was found to occur because the models were computed using
the gravity darkened equatorial temperature $T_{\rm eq}$ as
the argument of equation~(\ref{eq:Ne0}).
The most rapidly rotating stars with the lowest values of
$T_{\rm eq}$ were thus given the lowest disk densities.
A contributing factor was that these stars tended to be
late-type Be stars with the lowest effective temperatures and
the smallest radii.
The stellar radius comes into play because the electron-scattering
emissivity depends on the total number of electrons in the
emitting volume (i.e., on the product $N_{e} R_{\rm eq}$).
It remains to be shown whether the Be stars undergoing the most
rapid rotation indeed have lower-density disks as predicted above.
This seems somewhat counterintuitive, since whatever
mechanisms that produce the disks are likely to grow in
strength as the stellar rotation increases above the
disk formation threshold (presumably at $V_{\rm min}$).
However, for some combinations of stellar properties and
rotation rate, the mass loss from the disk may come to
dominate the angular momentum transport that feeds the disk,
and lower densities may occur because of greater leakage
to an equatorial wind.
More work needs to be done to compare the derived disk
properties of stars with similar spectral types but
different rotation rates.

\section{Further Evidence for Subcritical Be-star Rotation}

The results obtained above imply that early-type Be stars
can be rotating at significantly subcritical rates (down
to roughly half of their critical rotation speed).
This conclusion seems at odds with the recent statistical
arguments of Townsend et al.\  (2004) that suggest nearly
all Be stars could be rotating nearly critically.
There are several additional pieces of evidence that can
be brought to bear on this disagreement, and these ideas also
imply subcritical rotation for at least some Be stars.

Recent interferometric observations have put constraints
on the inclination angles of some bright Be stars.
Specifically, both Quirrenbach et al.\  (1997) and
Tycner et al.\  (2004) measured an extreme axial ratio for the
circumstellar envelope of $\zeta$~Tau that implies
$i > {70\arcdeg}$.
Using values of $v \sin i$ and $V_{\rm crit}$ from Slettebak
(1982) and Porter (1996), as well as assuming $i = {70\arcdeg}$
in order to produce an upper limit, Tycner et al.\  (2005)
estimated that $V_{\rm eq}/V_{\rm crit}$ for this star is
no more than $\sim$0.52.
Using the range of $V_{\rm crit}$ values computed in this
paper for $\zeta$~Tau and similarly dividing
$v \sin i/V_{\rm crit}$ by $\sin {70\arcdeg}$ yields a
slightly wider range of ratios: 0.48 to 0.56.
Assuming that gravity darkening was not taken into
account in the derivation of $v \sin i$, the strongest
possible correction would be to divide the above ratios by
the {\em smallest} allowable value of $D$ (i.e., 0.68 for a
B9 star), thus giving a range of $V_{\rm eq}/V_{\rm crit}$
for this star of 0.70--0.82.
Given that the minimum value of $D$ that corresponds to a B1
type star would be higher ($\sim$0.80), the subcritical
nature of this star's rotation appears definite.
Future interferometric measurements are expected to bring
about similar examples of firm subcritical rotation.\footnote{%
Several other stars discussed by Tycner et al.\  (2005)
were found to have ratios $V_{\rm eq}/V_{\rm crit}$ between
0.7 and 0.9, but applying the maximum possible correction for
gravity darkening brings these stars to nearly critical rotation.
The case of $\eta$~Tau, for which Tycner et al.\  give
a ratio of 0.53, seems to be anomalous because a main-sequence
$V_{\rm crit}$ was used for this giant star; using the
even the high-end value computed in this paper results in a
ratio $V_{\rm eq}/V_{\rm crit}$ of 0.80.
This star is thus in the same category as the five others
discussed by Tycner et al.\  (2005), excluding $\zeta$~Tau,
and more detailed analysis must be performed to determine the
star-specific corrections for gravity darkening.}

In addition to rotationally broadened spectral lines, another
potentially useful measure of near-critical rotation may be
the large-scale shape of a star's spectral energy distribution
(i.e., rotational ``color effects'').
It has been known for some time (e.g., Collins 1965;
Kodaira \& Hoekstra 1979) that gravity darkening can lead to
more pronounced color variations in the ultraviolet than in the
visible, though it is unclear how tightly these variations can
be calibrated to provide a direct measurement of the rotation
rate or the inclination.
Stalio et al.\  (1987) examined far-ultraviolet flux
distributions of several Be stars and was able to rule out
extreme gravity darkening signatures.
Stalio et al.\  made the preliminary conclusion that
$V_{\rm eq}/V_{\rm crit} \lesssim 0.85$ for the observed stars.

Collins \& Sonneborn (1977) argued that critical
rotation is unlikely for all Be stars because of the large
predicted photometric shift above the main sequence.
B-type stars rotating at 90\% of their critical angular
velocity (i.e., $V_{\rm eq}/V_{\rm crit} \approx 0.73$)
showed roughly a 1 magnitude increase in $V$-band absolute
magnitude, while critical rotators showed more than a
2 magnitude increase.
The presence of many Be stars at only $\sim$1 magnitude above
the main sequence seemed to Collins \& Sonneborn (1977) to be
evidence against critical rotation.
Note, though, that more recent calculations of photometric
shifts due to rapid rotation have predicted a substantially
weaker enhancement (see Figure 3 of Townsend et al.\  2004).

Mennickent et al.\  (1994) studied the detailed statistics of
H$\alpha$ spectral line shapes as a function of $v \sin i$.
From the observed spread of the measurements, they concluded
that there is ``...evidence for a considerable range of the
true rotation velocities of Be stars: definitely there are
intrinsically slow rotators among them.''

Finally, the observed {\em variability} of Be-star circumstellar
envelopes may be related to the division found in this paper
between subcritical rotation (for early-type Be stars) and
nearly critical rotation (for late-type Be stars).
Early-type Be stars are often observed to undergo strong
outburst phases with an inferred rapid evacuation and
refilling of the circumstellar disk region (see, e.g.,
Rivinius et al.\  1998, 2001).
Late-type Be stars are found to be comparatively quiescent,
with the transitions between Be and normal-B phases
being more gradual.
If the rotation rates of the early-type Be stars are far
below their respective $V_{\rm crit}$ limits, they may
require an impulsive mechanism of disk formation.
Models of isotropic ejection from a point on the star---with
some material being propelled forward into orbit and some
propelled backward to fall back onto the star---may be needed
to explain the disks around the hottest Be stars
(Kroll \& Hanuschik 1997; Owocki \& Cranmer 2002).
If, on the other hand, the coolest Be stars have nearly
critical rotation rates, rapid outbursts as described above
may not be needed for material to leak outwards into a
viscous decretion disk (see also Clark et al.\  2003;
Owocki 2005).

\section{Summary and Discussion}

This paper contains a statistical analysis of the equatorial
rotation rates of Be stars in Yudin's (2001) database.
A new Monte Carlo forward modeling procedure was developed to
simulate number distributions of $v \sin i$ for samples of
strongly gravity-darkened hot stars.
The parameters of the $V_{\rm eq}$ distributions that best
fit the observations were determined by a rigorous search of
parameter space and a minimization of the appropriate $\chi^2$
statistic.
Although there were some variations in the resulting parameters
depending on the assumed calibration of $V_{\rm crit}$,
the overall results seem robust.
Early-type (O7e--B2e) Be stars were found to exhibit a spread of
equatorial rotation speed between a lower limit of
0.4--0.6 $V_{\rm crit}$ and an upper limit of critical rotation.
Late-type (B3e--A0e) Be stars exhibit progressively narrower
ranges of rotation speed as $T_{\rm eff}$ decreases; the lower
limit rises gradually to 100\% of critical rotation
and the upper limit remains near the critical level.
Uncertainties in $V_{\rm crit}$ make it impossible to rule out
critical rotation for Be stars later than about $\sim$B3,
but the result of substantially {\em subcritical} rotation
for B0e--B2e stars appears to be firm.
This paper (see Figure 8) thus seems to imply the existence of a
rough dividing line in effective temperature between two
qualitatively different types of Be phenomenon:
\begin{enumerate}
\item
For $T_{\rm eff} \gtrsim 21000 \, K$ the threshold ratio
$V_{\rm min}/V_{\rm crit}$ above which a disk can form is well
below unity.
This seems to correspond to the regime of early-type Be stars
that undergo violent (possibly pulsation-driven) outbursts that
feed the circumstellar disk.
\item
For $T_{\rm eff} \lesssim 21000 \, K$ the threshold ratio
$V_{\rm min}/V_{\rm crit}$ ratio grows to unity as $T_{\rm eff}$
decreases to the end of the B spectral range.
The late-type Be stars at these temperatures are seen to be
more quiescent than their early-type counterparts, and this
may be due to the relative ease of generating a Keplerian disk
for their closer proximity to critical rotation.
\end{enumerate}
The above results for early-type Be stars are in disagreement
with other recent suggestions that the Be phenomenon is closely
linked to critical or near-critical rotation (see, e.g.,
Maeder \& Meynet 2000b).
Pushing the $V_{\rm min}$ threshold down to significantly
subcritical rotation speeds makes it more difficult to explain
the existence of Keplerian decretion disks around Be stars.

The analysis presented in this paper can be improved in several
ways to increase confidence in the results.
The observational determination of $v \sin i$, $V_{\rm crit}$,
and possibly also the inclination angle $i$ itself can be
done more rigorously for an individual star when more than just
one line profile is available (Stoeckley \& Buscombe 1987;
Reiners 2003; Fr\'{e}mat et al.\  2005).
For large $v \sin i$ databases like that of Yudin (2001), the
tabulated observational uncertainties should be utilized to a
greater degree in producing comparable simulated samples.
The assumption used in this paper of normally distributed
uncertainties normalized by a single value of $\sigma_{\zeta}$
may have contributed to systematic biases that a more
observationally guided procedure could correct.

The various analyses performed here should also be done for
catalogs of normal B-type stars that do not exhibit emission
lines (see, e.g., G{\l}\c{e}bocki \& Stawikowski 2000;
Abt et al.\  2002).
It is still unclear if exceeding the minimum threshold
rotation speed determined in this paper is both necessary and
sufficient for the onset of the Be phenomenon; there may be
many non-Be stars with $V_{\rm eq} > V_{\rm min}$.
It is also not clear what sets the observed fraction of
Be-star incidence as a function of the larger population of
early-type stars (see, e.g., Zorec \& Briot 1997;
Penny et al.\  2004).
Firmer constraints on the rotation distributions of each
population would be helpful.
The definition of a truly ``normal'' B star is problematic,
though, because a Be star can spend many years in a phase
without a circumstellar disk and thus would have exhibited no
Balmer emission over the entire time period of observations.
There may be many early-type stars that will eventually become
Be stars but are still classified as normal B stars.

Our understanding of the rotation threshold for the onset of
the Be phenomenon can be aided by incorporating other kinds of
observations that were not studied in this paper.
Better interferometric measurements that can constrain tightly
both the stellar oblateness and the circumstellar disk geometry
(McAlister et al.\  2005; Tycner et al.\  2005) are becoming
more widely available.
In the long term, space-based missions like the proposed
{\em Stellar Imager} may provide direct images of the oblate
and gravity darkened stellar surfaces as well as firm
constraints on differential rotation and macroturbulence
(Carpenter et al.\  2004).
Measuring the widths of lines that are formed in the narrow
``boundary layer'' above the subcritical photosphere, but below
the inner edge of the Keplerian disk (e.g., Chen et al.\  1989),
may be a crucial probe of the physical origin of the
Be phenomenon.

\acknowledgments

I gratefully acknowledge Adriaan van Ballegooijen,
Ian Howarth, Richard Townsend, and Stan Owocki
for many valuable discussions,
as well as Ruslan Yudin, Antonio Claret Dos Santos, and
Bob Kurucz for making available their useful databases.
This work was supported by the National Aeronautics and Space
Administration (NASA) under grant {NNG\-04\-GE77G} to the
Smithsonian Astrophysical Observatory.
This research made extensive use of NASA's Astrophysics
Data System.

\clearpage
\small

\begin{deluxetable}{lccccc}

\tablecaption{Be Star Rotational Properties by Subpopulation}
\tablewidth{0pt}

\tablehead{
\colhead{ }  &
\colhead{VE} & \colhead{E}  & \colhead{M}  & 
\colhead{L}  & \colhead{VL}
}

\startdata

\sidehead{Definitions:}

min $(T_{\rm eff})$ (K) &
  24000  &  20500  &  18500  &  13200  &  10000  \\
max $(T_{\rm eff})$ (K) &
  36000  &  24000  &  20500  &  18500  &  13200  \\
approx.\  spectral types &
  O7--B1.5  &  B1.5--B2.5  &  B2.5--B3.5  &  B3.5--B6  &  B6--A0  \\
number of stars &
  93  &  93  &  92  &  92  &  92  \\

\sidehead{Observed statistics (Cranmer's $V_{\rm crit}$):}

mean $(v \sin i / V_{\rm crit})$ &
 {\phs}0.4069 & {\phs}0.4335 & {\phs}0.4740 & {\phs}0.5269 & {\phs}0.5863 \\
std.\  dev.\  $(v \sin i / V_{\rm crit})$ &
 {\phs}0.1527 & {\phs}0.1401 & {\phs}0.1693 & {\phs}0.1619 & {\phs}0.1490 \\
skewness $(v \sin i / V_{\rm crit})$ &
 {\phs}0.2625 & {\phs}0.3260 & $-$0.1048    & $-$0.4354    & $-$0.1232    \\
kurtosis $(v \sin i / V_{\rm crit})$ &
 {\phs}0.1242 & $-$0.4883    & $-$0.1489    & $-$0.5136    & $-$0.3359    \\

\sidehead{Observed statistics (Chauville et al.\  $V_{\rm crit}$):}

mean $(v \sin i / V_{\rm crit})$ &
 {\phs}0.4926 & {\phs}0.5225 & {\phs}0.5812 & {\phs}0.6795 & {\phs}0.8957 \\
std.\  dev.\  $(v \sin i / V_{\rm crit})$ &
 {\phs}0.1837 & {\phs}0.1661 & {\phs}0.2061 & {\phs}0.2144 & {\phs}0.2544 \\
skewness $(v \sin i / V_{\rm crit})$ &
 {\phs}0.2230 & {\phs}0.1964 & $-$0.4215    & $-$0.7592    & $-$1.2430    \\
kurtosis $(v \sin i / V_{\rm crit})$ &
 {\phs}0.3961 & $-$0.7920    & $-$0.5134    & $-$0.3913    & $-$0.4913    \\

\sidehead{Best-fit $V_{\rm eq}$ distributions
(Cranmer's $V_{\rm crit}$):}

$V_{\rm min}/V_{\rm crit}$ &
 {\phs}0.4150 & {\phs}0.3423 & {\phs}0.5050 & {\phs}0.7219 & {\phs}0.9084 \\
$V_{\rm max}/V_{\rm crit}$ &
 {\phs}0.7862 & {\phs}0.7955 & {\phs}0.9118 & {\phs}0.7454 & {\phs}0.9121 \\
$S$ &
 $-$0.9500 & {\phs}0.0886 & $-$0.7645 & {\phs}0.6904 &  0 \\
min $(\chi^{2} / \nu)$ &
 0.39  &  1.31  &  0.58  &  0.44  &  0.57  \\
$Q (\chi^{2} | \nu)$, in \% &
 95.3  &  21.6  &  83.5  &  92.5  &  83.6  \\
$\langle V_{\rm eq} / V_{\rm crit} \rangle$ &
 {\phs}0.5416 & {\phs}0.5756 & {\phs}0.6562 & {\phs}0.7361 & {\phs}0.9103 \\
best $\sigma_{\zeta}$ &
 0.20 & 0.15 & 0.20 & 0.15 & 0.15 \\

\sidehead{Best-fit $V_{\rm eq}$ distributions
(Chauville et al.\  $V_{\rm crit}$):}

$V_{\rm min}/V_{\rm crit}$ &
 {\phs}0.4914 & {\phs}0.3573 & {\phs}0.8832 & $\gtrsim 0.95$ &
 $\gtrsim 0.95$ \\
$V_{\rm max}/V_{\rm crit}$ &
 {\phs}0.9377 & {\phs}0.9999 & {\phs}0.8886 & --- & --- \\
$S$ &
 $-$1 &  {\phs}0.5882  & 0 & --- & --- \\
min $(\chi^{2} / \nu)$ &
 0.47  &  0.99  &  1.15  &  ---   &  ---   \\
$Q (\chi^{2} | \nu)$, in \% &
 91.4  &  45.1  &  32.0  &  ---   &  ---   \\
$\langle V_{\rm eq} / V_{\rm crit} \rangle$ &
 {\phs}0.6404 & {\phs}0.7419 & {\phs}0.8859 & --- & --- \\
best $\sigma_{\zeta}$ &
 0.20 & 0.15 & 0.20 & ---  & ---  \\

\enddata

\end{deluxetable}

\clearpage
 
\begin{figure}
\epsscale{0.70}
\plotone{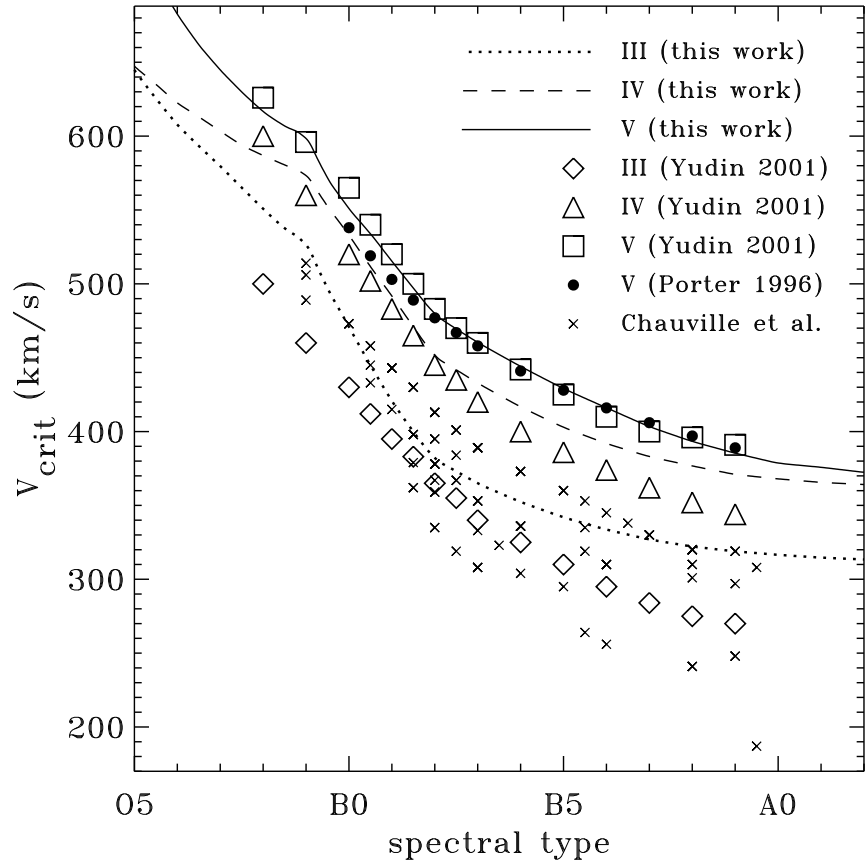}
\caption{
Critical rotation speed as a function of spectral type for
stars of luminosity class III ({\em dotted line}),
IV ({\em dashed line}), and V ({\em solid line}).
Symbols denote values of $V_{\rm crit}$ taken from other
papers (see labels above).}
\end{figure}

\clearpage
 
\begin{figure}
\epsscale{0.70}
\plotone{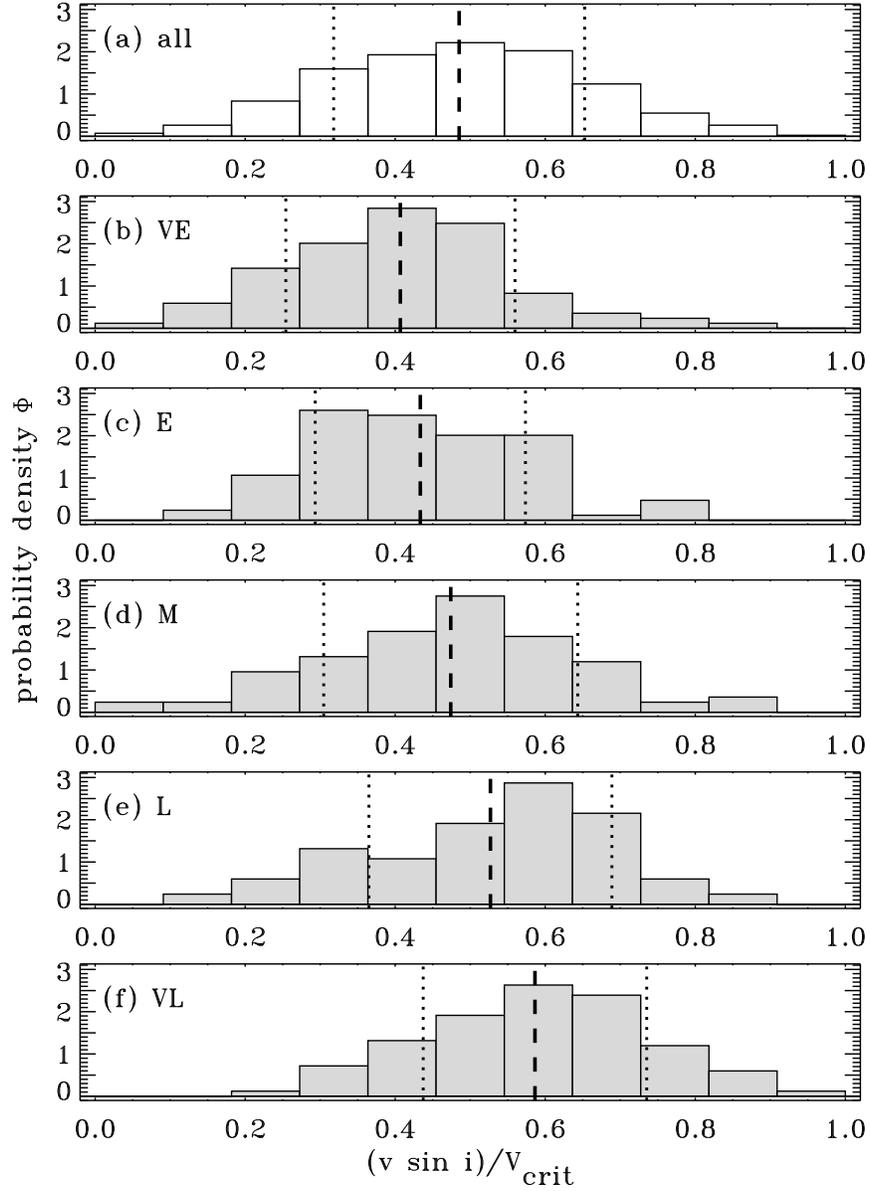}
\caption{
Observed number distributions of $v \sin i / V_{\rm crit}$
for: (a) the full Yudin (2001) database of 462 stars having
nonzero $v \sin i$; (b)--(f) the five subpopulations,
defined by the $T_{\rm eff}$ bounds given in in Table 1.
These ratios are computed using the high-end (Cranmer)
calibration for $V_{\rm crit}$.
Also shown are the means ({\em{dashed lines}}) and
$\pm$1$\sigma$ standard deviations ({\em{dotted lines}})
for each of the independently normalized distributions.}
\end{figure}

\clearpage
 
\begin{figure}
\epsscale{0.45}
\plotone{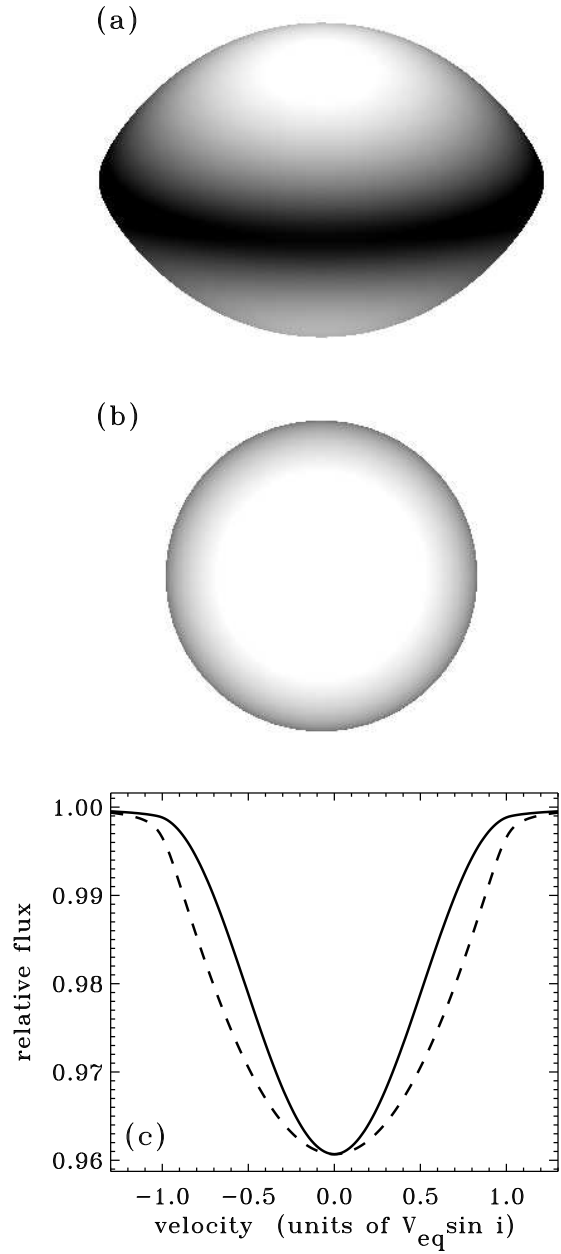}
\caption{
Schematic continuum intensity images of a B9 V star rotating
rigidly with $V_{\rm eq} = 0.95 \, V_{\rm crit}$ and viewed
from an inclination angle $i = 75\arcdeg$ measured from the pole.
Both images are derived with the same linear limb darkening, but
they differ in that (a) uses the centrifugal force to alter the
surface equipotential surfaces (Roche oblateness) and
redistributes the radiative flux in proportion to the effective
gravity (von Zeipel gravity darkening), and (b) does not.
Panel (c) shows resulting absorption lines computed as described
in the text: the solid curve corresponds to image (a); the dashed
curve corresponds to (b).  The central depth of profile (b) was
renormalized to match that of (a) to more clearly compare the
line shapes.}
\end{figure}

\clearpage
 
\begin{figure}
\epsscale{0.70}
\plotone{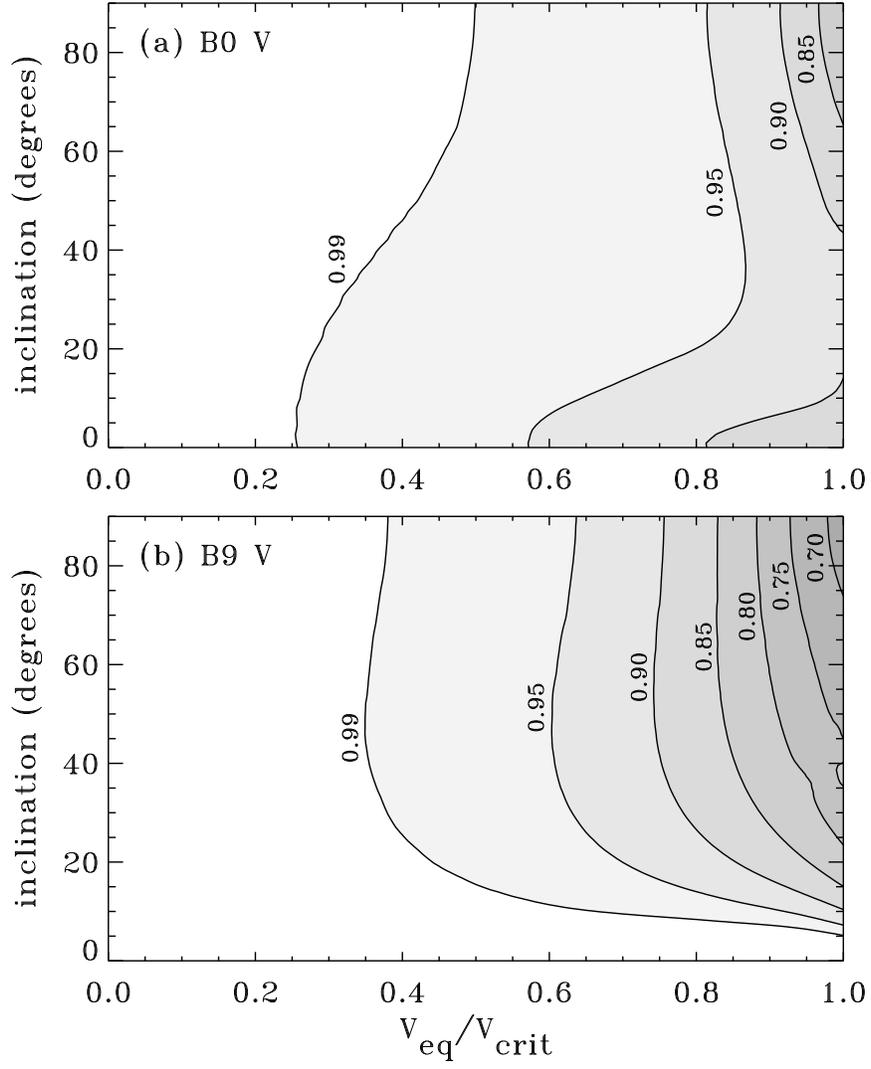}
\caption{
Contour plots of the line narrowing ratio $D$ due to
gravity darkening, for main sequence B0 (a) and B9 (b) models.
The values of the constant levels are labeled
above (white denotes $D \approx 1$).}
\end{figure}

\clearpage
 
\begin{figure}
\epsscale{0.70}
\plotone{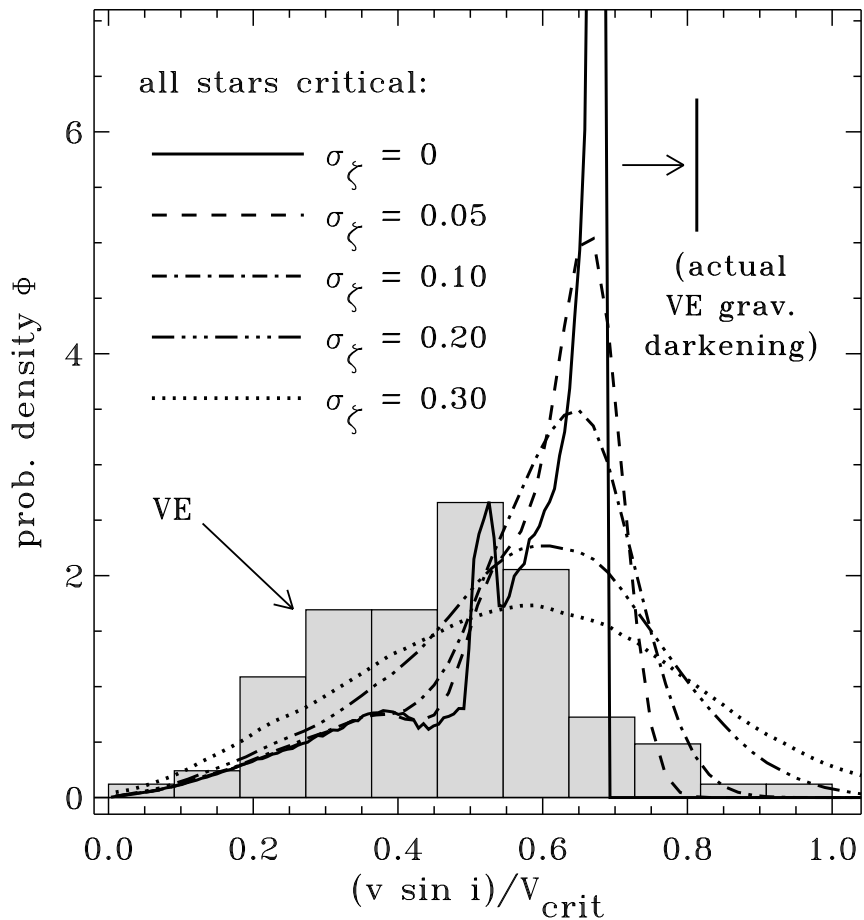}
\caption{
Statistical distributions $\Phi_{j}(v \sin i)$ computed under
the assumption that all stars are rotating at $V_{\rm crit}$
and using the strongest (i.e., B9) gravity darkening grid for $D$.
The simulated observations were made with varying levels of
random uncertainty: $\sigma_{\zeta} = 0$ ({\em{solid line}}),
0.05 ({\em{dashed line}}), 0.1 ({\em{dot-dashed line}}), 0.2
({\em{dash-triple-dot line}}), and 0.3 ({\em{dotted line}}).
Also plotted ({\em{gray bars}}) is the observed distribution
of $v \sin i / V_{\rm crit}$ values for the 93 stars in the
VE subpopulation, normalized using the fit to the
Chauville et al.\  (2001) low-end $V_{\rm crit}$ values.
A vertical line in the upper-right shows how the distributions
would be shifted to higher values of $v \sin i$ if the
consistent gravity darkening grid for the VE stars (i.e., B0)
would have been used.}
\end{figure}

\clearpage
 
\begin{figure}
\epsscale{1.00}
\plotone{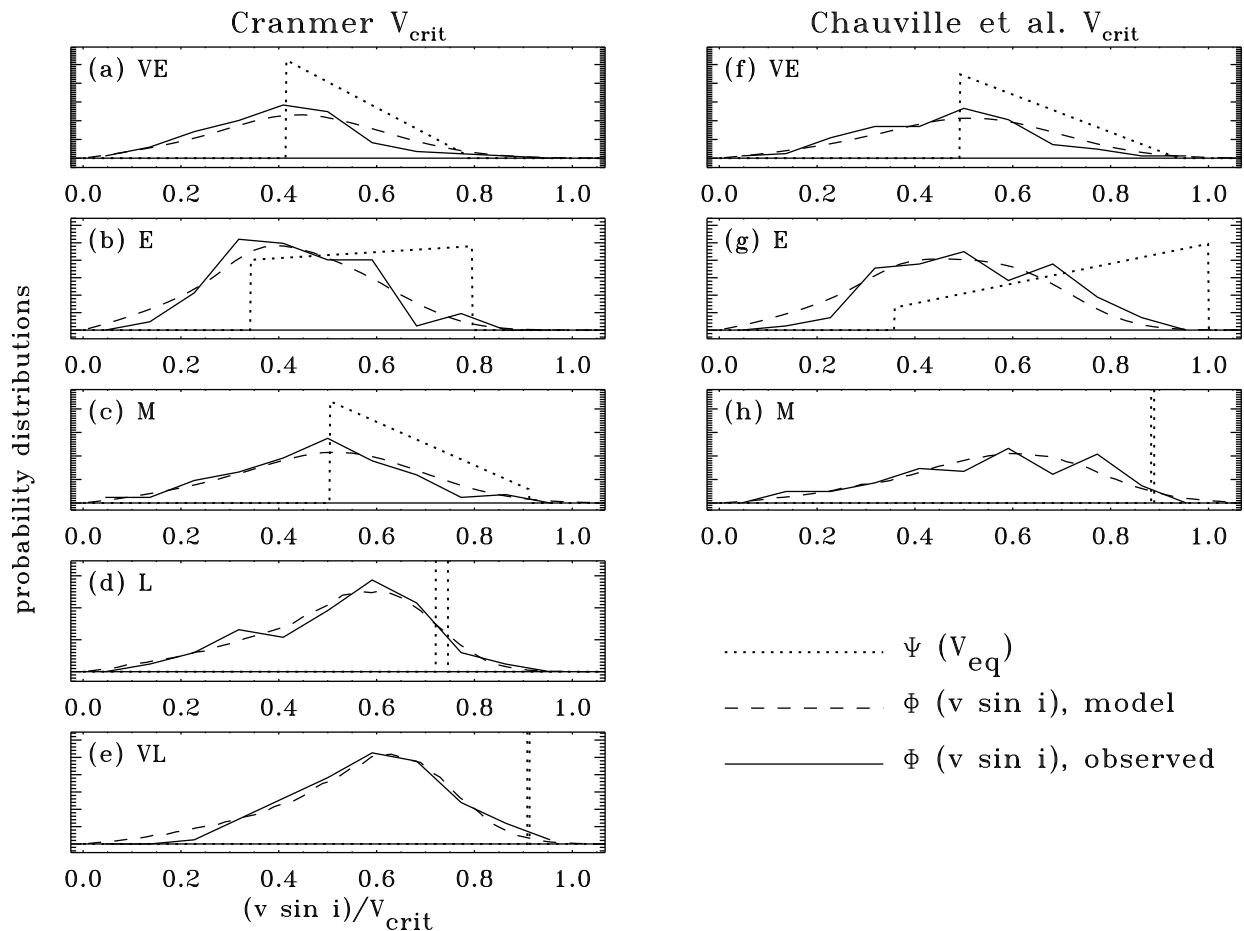}
\caption{
Comparisons of observed ({\em{solid lines}}) and simulated
({\em{dashed lines}}) probability distributions of the
projected rotation speed $\Phi (v \sin i)$, as well as the
best-fit distributions of intrinsic rotation speed
$\Psi (V_{\rm eq})$ ({\em{dotted lines}}) for the 5
spectral subpopulations defined in Table 1.
(a)--(e): calculations assuming the high-end (Cranmer)
calibration for $V_{\rm crit}$ derived in {\S}~2.
(f)--(h): calculations assuming the low-end (Chauville et al.)
calibration for $V_{\rm crit}$.
The L and VL models for the latter calibration did not
yield acceptable solutions and are not shown.}
\end{figure}

\clearpage
 
\begin{figure}
\epsscale{0.70}
\plotone{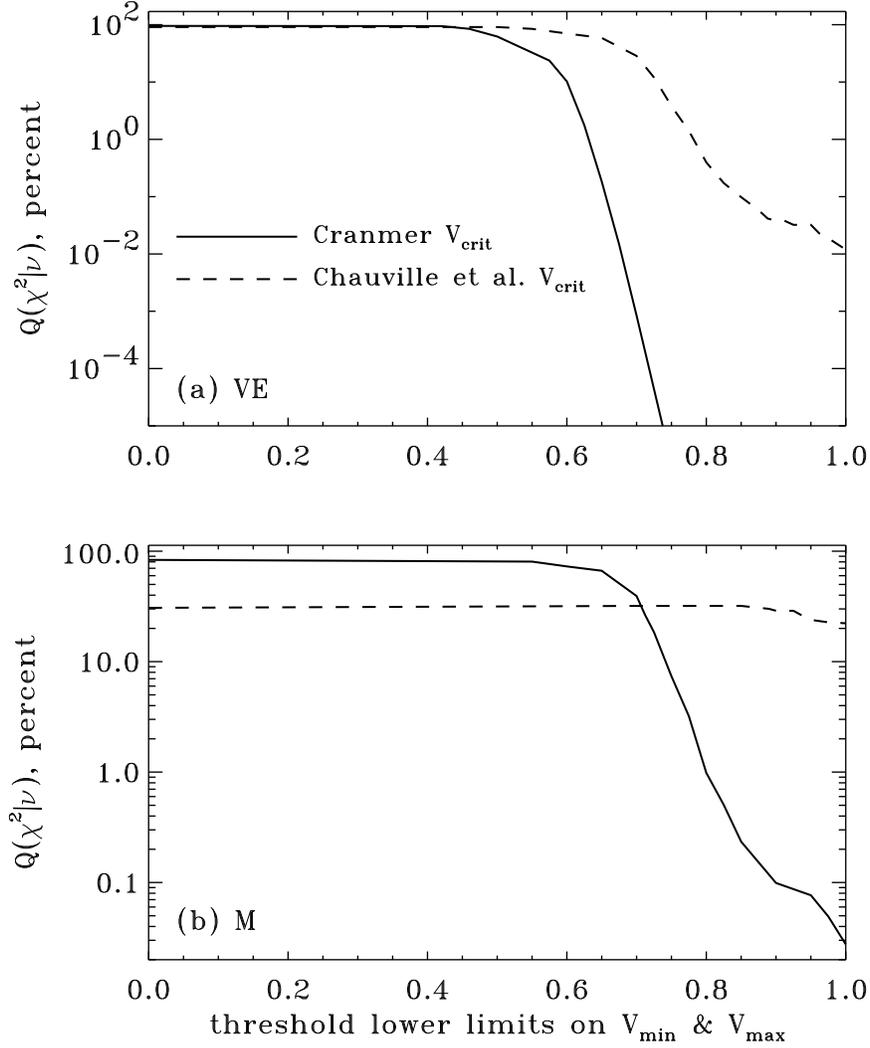}
\caption{
Threshold goodness-of-fit probabilities $Q$ for the (a) VE
and (b) M spectral subranges, computed for a range of
constrained searches of parameter space.
Each data point represents the maximum probability allowed
when $V_{\rm min}$ and $V_{\rm max}$ were required to exceed
the threshold values plotted as the abscissa.
Parameter searches were performed assuming both the
high-end (Cranmer) $V_{\rm crit}$ calibration
({\em{solid lines}}) and the low-end (Chauville et al.)
$V_{\rm crit}$ calibration ({\em{dashed lines}}), both with
$\sigma_{\zeta}$ kept constant at 0.20.}
\end{figure}

\clearpage
 
\begin{figure}
\epsscale{0.70}
\plotone{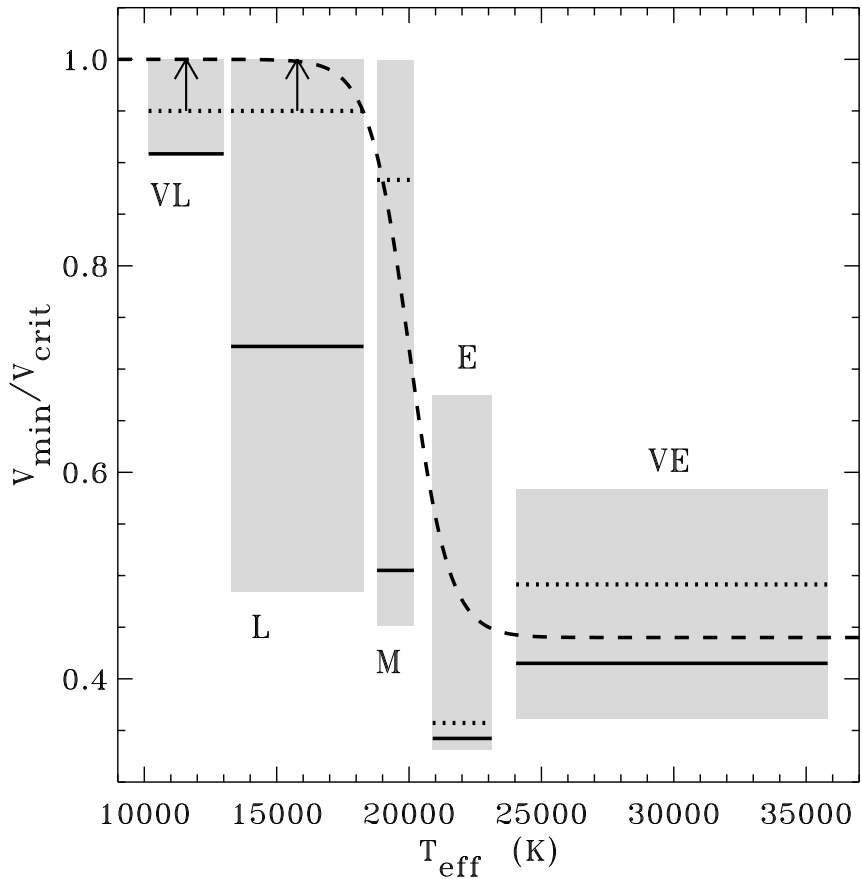}
\caption{
Minimum allowed equatorial rotation speeds for the observed
populations of Be stars, plotted as dimensionless ratios
$V_{\rm min}/V_{\rm crit}$ versus $T_{\rm eff}$.
The empirical limits determined from $\chi^{2}$ minimization
(see Table 1) are plotted as horizontal bars corresponding to the
high-end Cranmer $V_{\rm crit}$ calibration ({\em{solid lines}})
and the low-end Chauville et al.\   $V_{\rm crit}$ calibration
({\em{dotted lines}}).
The gray bars show the range of $V_{\rm min}$ values that occur
when $\sigma_{\zeta}$ is varied between 0 and 0.30 in steps of
0.05, and only high-probability values with
$Q (\chi^{2} | \nu) > 10$\% are kept.
Also shown is the approximate fitting formula given by
equation~{\ref{eq:Veqmin}} ({\em{dashed line}}).}
\end{figure}

\clearpage
 
\begin{figure}
\epsscale{0.70}
\plotone{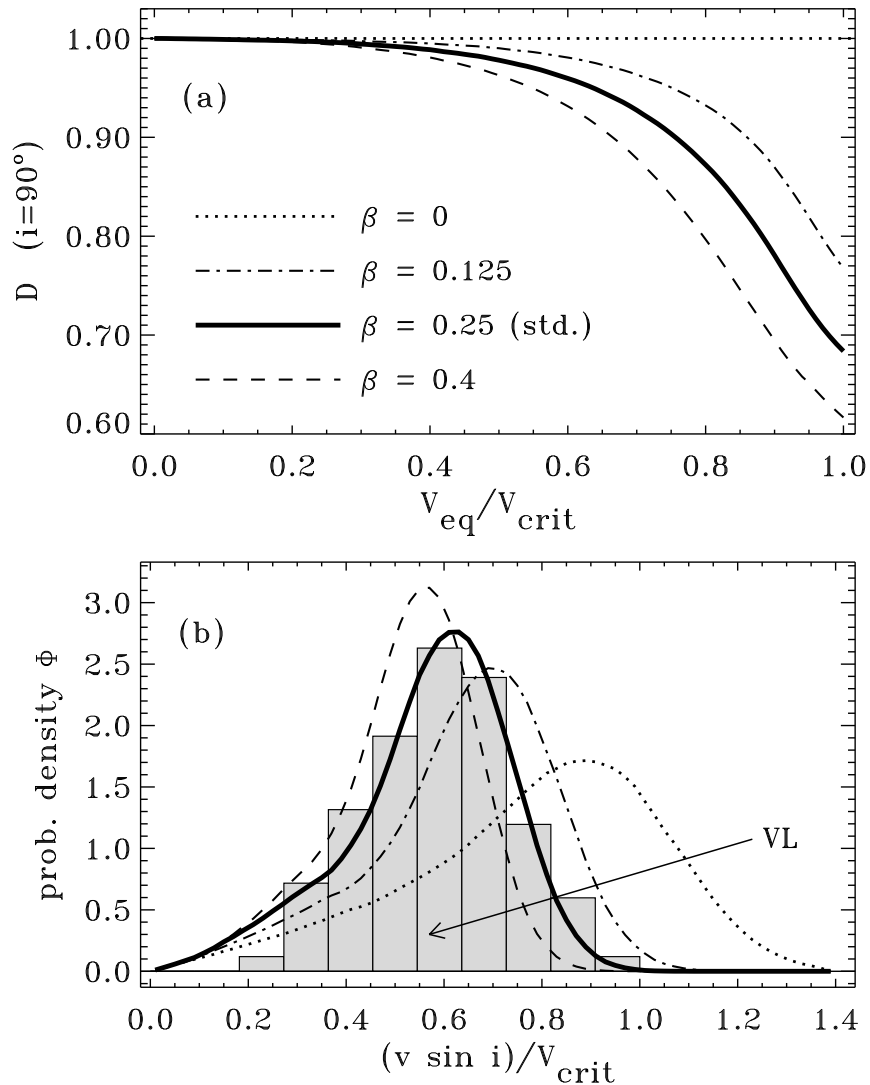}
\caption{
(a) Line narrowing ratio $D$ versus equatorial rotation speed
for a constant inclination angle of {90\arcdeg} and a range
of gravity darkening exponents $\beta$ (see labels above).
(b) Statistical distributions $\Phi_{j}(v \sin i)$ computed
for the same range of $\beta$ values as in panel (a),
assuming that all stars are rotating at $V_{\rm crit}$ and
have a mean uncertainty level $\sigma_{\zeta} = 0.15$.
Also plotted ({\em{gray bars}}) is the observed distribution
of $v \sin i / V_{\rm crit}$ values for the 92 stars in the
VL subpopulation normalized by the high-end Cranmer calibration
for $V_{\rm crit}$ (see also Fig.~2f).}
\end{figure}

\clearpage
 
\begin{figure}
\epsscale{0.70}
\plotone{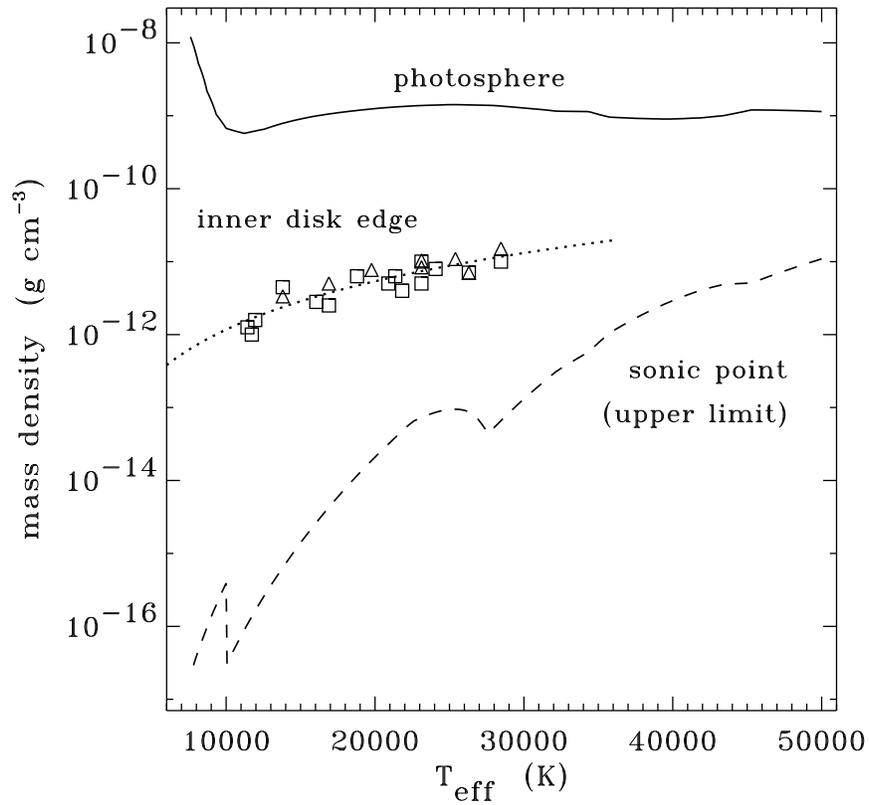}
\caption{
Mass densities (in g cm$^{-3}$) computed for main sequence
nonrotating B stars at various heights, versus stellar
effective temperature:
optical-depth-unity photosphere ({\em{solid line}}),
an upper limit estimate for the equatorial sonic point of disk
outflow ({\em{dashed line}}), and the fitting formula for
the inner disk density given in equation~(\ref{eq:Ne0})
({\em{dotted line}}).
Also shown are 8 measurements of inner disk densities from
Table 2 of McDavid 2001 ({\em{triangles}}) and 15 measurements
from Tables 3 and 5a of Waters et al.\  1987 ({\em{squares}}).}
\end{figure}

\clearpage
 
\begin{figure}
\epsscale{0.70}
\plotone{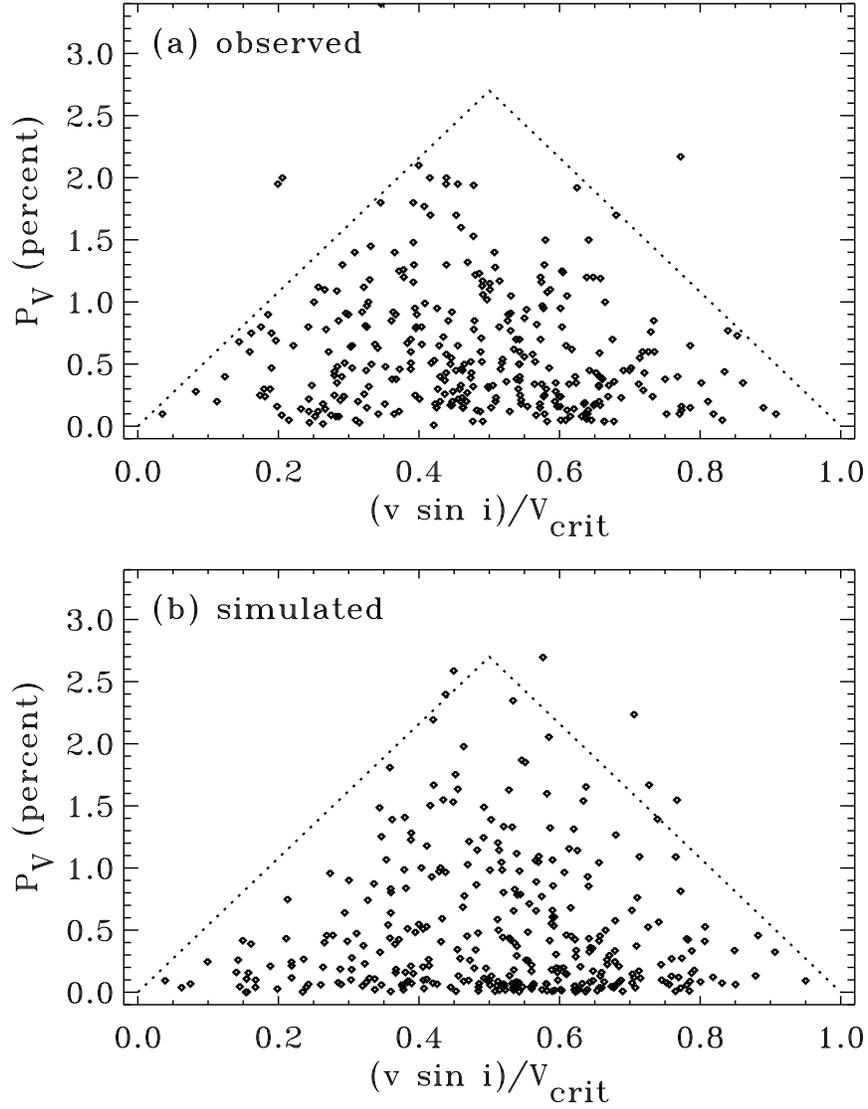}
\caption{
Correlation between visible-band polarization $P_V$ and
projected rotation speed $v \sin i$, shown for:
(a) 335 observed stars with nonzero measurements for both
values from the Yudin (2001) database, and
(b) 335 simulated stars, using a modified version of the
method of McDavid (2001).
Rough upper bounds to the spread of $P_V$ values are also
plotted ({\em{dotted lines}}).}
\end{figure}

\end{document}